\documentclass[a4paper,11pt]{article}
\pdfoutput=1 

\usepackage{jheppub} 

\usepackage[T1]{fontenc} 

\title{\boldmath Type IIB \textit{parabolic} ($p,q$)-strings from M2-branes with fluxes.}

\author{M.P. García del Moral,}
\author{C. las Heras}
\author{and A. Restuccia\footnote{The authors were listed in alphabetical order.}}
\affiliation{Departamento de Física, Universidad de Antofagasta,\\Antofagasta, Chile.}

\emailAdd{maria.garciadelmoral@uantof.cl}
\emailAdd{camilo.lasheras@ua.cl}
\emailAdd{alvaro.restuccia@uantof.cl}

\abstract{We extend the work of Schwarz \cite{Schwarz6} to show that bound states of type IIB supersymmetric ($p, q$)-strings on a circle are associated with M2-branes irreducibly wrapped on $T^2$, or equivalently with nontrivial worldvolume fluxes. Beyond this extension we consider the Hamiltonian of an M2-brane with $C_{\pm}$ fluxes formulated on a symplectic torus bundle with monodromy. In particular, we analyze the relevant case when the monodromy is parabolic. We show that the Hamiltonian is defined in terms of the coinvariant module. We also find that the mass operator is invariant under transformations between inequivalent coinvariants. These coinvariants classify the inequivalent classes of twisted torus bundles with nontrivial monodromy for a given flux. We obtain their associated ($p,q$)-strings via double dimensional reduction, which are invariant under a parabolic subgroup of $SL(2,\mathbb{Q})$. This is the origin of the gauge symmetry of the associated gauged supergravity. These bound states could also be related to the parabolic Scherk-Schwarz reductions of type IIB string theory.
}

\begin{document} 
\maketitle
\flushbottom

\section{Introduction}
Type IIB supersymmetric theories have an $SL(2,Z)$ invariance that allows to define bound states of $(p,q)$-strings. These can be interpreted as nonperturbative states of $p$ fundamental strings ($F1$) with $q$ D$1$-branes \cite{Schwarz6,Witten4}.  Bound states of strings and/or D-branes have been studied in different scenarios, as for example, in terms of black holes \cite{Strominger2,Berenstein}, or to define boundary states in the context of AdS/CFT duality \cite{Filippas}, see also \cite{Roy2,Alexandrov,Alexandrov2}. Bound states of ($p,q$)-strings in a background with fluxes were studied in \cite{Kluson}. They showed that ($p,q$)-strings on a $AdS_3\times S^3$ background with mixed (RR and NSNS) three-form fluxes are mapped into $(p',q')$-strings on the same background with NSNS three-form fluxes by means of an $SL(2,Z)$ transformation.

An $SL(2,Z)$ family of type IIB ($p,q$)-string solutions were obtained in \cite{Schwarz6,Schwarz7}, with $p$ and $q$ coprime representing the NSNS and RR charges, respectively. Its supersymmetric extension was obtained from M2-branes toroidally wrapped on general backgrounds in \cite{Uehara}. The well-defined action of T-duality on D$p$-branes \cite{Bergshoeff6}, allows to define bound states of fundamental strings with D$p$-branes \cite{Roy}. The action of S-duality on bound states of ($p,q$)-strings is also well-known to mix the RR and NSNS charges. The T-duality between type II theories and the identification of type IIA with 11D supergravity on a circle induces the relation of type IIB on a circle with 11D supergravity on a torus. Because of this relationship, \cite{Schwarz6} proposed that the KK and winding terms of the mass operator of ($p,q$)-strings compactified on a circle are obtained from the mass operator of an M2-brane on a torus \cite{mpgm21}. 

The M2-brane on a flat space, more precisely the regularized $SU(N)$ formulation, has a continuous spectrum from $[0,+\infty)$ \cite{deWit6,deWit2}. For this reason M2-branes have not been considered as fundamental objects describing microscopic degrees of freedom of M-theory. This result led to the matrix theory conjecture \cite{Susskind}, see also \cite{Nicolai}, where M2-branes were interpreted as second quantized theory. However, in \cite{Restuccia} it was noticed that M2-branes compactified on a torus, have two well defined sectors related to the imposition of a topological restriction named ``central charge condition''. It implies the irreducible wrapping of the M2-brane on the compact sector, which ensures that the determinant of the winding matrix is nonzero. It has been rigorously proved that M2-brane with central charges has a discrete supersymmetric spectrum with finite multiplicity \cite{Boulton}. Therefore, the M2-brane with central charges describes the microscopical degrees of freedom of a well defined sector of M-theory. Furthermore, M2-branes with central charges are characterized by a 2-form flux condition on the worldvolume and they also contain a symplectic structure \cite{mpgm6}. The sector without central charges defined on the same target space, is related to a reducible wrapping on the compact sector, which corresponds to a winding matrix with a trivial determinant and a continuous supersymmetric spectrum.

The Hamiltonian formulation for an M2-brane in the light-cone gauge on $M_9\times T^2$, on a constant supergravity background with a $C_{\pm}$ flux condition on $T^2$, has a discrete supersymmetric spectrum, as shown in \cite{mpgm6}. In fact, the 2-form flux condition on the target space is in one-to-one correspondence with the central charge condition, and the Hamiltonian is related by a canonical transformation of the phase space variables. The flux condition implies that the M2-brane is wrapped irreducibly around the compact sector. It means that the determinant of the wrapping matrix is non zero. Consequently, M2-branes with $C_\pm$ fluxes are equivalent to M2-branes with central charges. It can be seen from \cite{mpgm10} that the symplectic structure and the flux condition present in both sectors are compatible. Indeed, the global description is given in terms of twisted torus bundles with monodromy in $SL(2,Z)$ and it contains nontrivial $U(1)$ gauge symmetries on the worldvolume \cite{mpgm10}.

In the original paper of Schwarz, the ($p,q$)-string mass operator on $M_9\times S^1$ was obtained from the mass operator of an ``M2-brane wrapped $n$ times on a torus" \cite{Schwarz6}. The KK and winding terms were obtained by Schwarz, however the terms corresponding to the oscillators were not computed. They were proposed as membrane excitations, but their expressions in terms of the geometric moduli and physical degrees of freedom were not obtained. Nevertheless, in  \cite{mpgm21}, the bosonic part of this contribution was explicitly computed and shown that it corresponds to the sector of M-theory named as M2-brane with central charges, which is characterized by a discrete supersymmetric spectrum. Indeed, the central charge condition ensures the appearance of the winding and KK terms on the M2-brane mass operator, as well as the computation of, at least, the bosonic sector of the terms corresponding to the oscillators. This is the origin of the oscillators on the type IIB ($p,q$)-string in $M_9\times S^1$.

The purpose of this paper is to obtain the string description of the well-behaved sectors of M2-brane on a torus. We obtain the ($p,q$)-string mass operator on $M_9\times S^1$ from an M2-brane with nontrivial fluxes and nontrivial monodromies, not considered in \cite{Schwarz6} or \cite{mpgm21}. It contains a new term, not obtained in \cite{mpgm21}, that depends on the flux contribution, the moduli of theory and the tension. Moreover, we perform a formulation that is completely supersymmetric in contrast with \cite{mpgm21} where only bosonic degrees of freedom were considered. We obtain a new class of ($p,q$)-strings from supermembranes with nontrivial monodromies, in distinction with the analysis done in \cite{Schwarz6} and \cite{mpgm21}, where only trivial monodromy was considered. In fact the formulation in \cite{Schwarz6,mpgm21} is a particular case of the general one in this work. The monodromy group is a subgroup of $SL(2,\mathbb{Z})$ (the group of isotopy classes of the area preserving diffeomorphisms, the latter is the structure group of the M2-brane). For a given monodromy group, the inequivalent formulations are classified by its coinvariants, which identify equivalent ($p,q$)-charges. The orbits generated by the monodromy subgroup acting on the elements ($p,q$) of a coinvariant are contained in the same coinvariant. A formulation in terms of the coinvariants induces the appeareance of a new symmetry, that is a parabolic subgroup of $SL(2,\mathbb{Q})$ for the case of parabolic monodromies. This is the M-theory origin of the parabolic gauging in type II gauged supergravity. Finally, we are generalizing the computation of the standard type IIB $(p,q)$-strings on a circle to the supersymmetric degrees of freedom. It contains a new term inherited from the flux condition. We are also emphasizing that it is directly related to the sectors with discrete supersymmetric spectrum in $M_9\times T^2$ and trivial monodromy. Nevertheless, at least for parabolic monodromies, we are identifying a particular type of ($p,q$)-string, which we have named "parabolic". The low energy limit of the corresponding theory of parabolic ($p,q$)-strings is not maximal but gauge supergravity.

The paper is organized as follows: In section 2, we briefly introduced the local and global descriptions of supermembranes with $C_\pm$ fluxes. In section 3, we show that the full mass operator of a ($p,q$)-string compactified on a circle is obtained from the  M2-brane with central charge. We show that the irreducible wrapping condition is necessary to obtain the ($p,q$)-string. In section 4 we discuss the M2-brane twisted torus bundle inequivalent theories. We obtain that the mass operator is invariant on an orbit of charges generated by the monodromy \cite{mpgm7}. In section 5, we show that for parabolic monodromies, M2-branes with fluxes are invariant on the classes of coinvariants. In this case, we also obtain that the mass operator is invariant under a transformation between the coinvariants that classify the second cohomology group of the bundle, hence, connecting inequivalent M2-brane bundles. In section 6, we obtain  a new type of  $(p,q)$-strings with restricted discrete symmetry given by the parabolic monodromy of the M2-brane twisted torus bundle that we denote under the name of `parabolic' ($p,q$)-string. In section 7, we present a brief discussion and our conclusions.

\section{M2-branes on twisted torus bundles.}
The light cone gauge (LCG) bosonic Hamiltonian for an M2-brane in the presence of a non-vanishing three-form background was given in \cite{deWit}, where the authors have considered a general spacetime with metric $G_{\mu\nu}$ written in a convenient form using the gauge $G_{--}=G_{a-}=0$. Its supersymmetric extension on a local Minkowski spacetime ($M_{11}$) was obtained in \cite{mpgm6} (where it was also shown its consistency with $C_{\mu\nu\rho}$ constant, not necessarily zero) and it is given by
\begin{eqnarray}\label{HCM2}
\mathcal{H}&=&\left[\frac{1}{(\widehat{P}_--TC_-)}\left(\frac{1}{2}(\widehat{P}_a-TC_a)^2+\frac{T^2}{4}(\epsilon^{uv}\partial_u X^a \partial_v X^b)^2\right) \right. \nonumber  \\
&-& \left. T\bar{\theta}\Gamma^-\Gamma_a \left\lbrace X^a,\theta \right\rbrace - TC_{+-}- TC_+ \right], 
\end{eqnarray}
subject to the first and second class constraints
\begin{eqnarray}
\widehat{P}_a\partial_u X^a + \widehat{P}_- \partial_u X^- + \bar{S}\partial_u \theta &\approx& 0, \\
S - (\widehat{P}_--TC_-)\Gamma^- \theta &\approx& 0 ,
\end{eqnarray}
with $T$ being the M2-brane tension and the unique free parameter of the theory, $\widehat{P}_a$ the canonical conjugate to $X^a$ and $S,\bar{S}$ are the conjugate momenta to $\bar{\theta},\theta$ (Majorana spinors in 11D), respectively \cite{deWit2}. 

The embedding used in this paper, is the same one used in \cite{deWit2,deWit6} seminal papers. We are considering the light cone gauge and hence the space-time indices $\mu,\nu,\rho=0,\dots,10$ are splitted according to $\mu=(+,-,a)$, where $a=1,\dots,9$ are the transverse indices to the null light coordinates (see for example \cite{deWit2}). The worldvolume indices are $i=0,1,2$, with $u,v=1,2$ labeling the spatial coordinates. We are considering an embedding of the M2-brane on the complete 11D space-time. That is, $X^a(\sigma^1,\sigma^2,\tau)$ are maps from $\Sigma$, a Riemann surface of genus 1, to the target space, $X^a : \Sigma \rightarrow M_{11}$.

The LCG three-form components are written according to \cite{deWit} as
\begin{equation}\label{CaLCG}
\small
\begin{aligned}
& C_a  =  -\epsilon^{uv}\partial_uX^- \partial_vX^b C_{-ab} +\frac{1}{2}\epsilon^{uv}\partial_uX^b \partial_vX^c C_{abc} \, , \\
& C_{\pm}  =  \frac{1}{2}\epsilon^{uv}\partial_uX^a \partial_vX^b C_{\pm ab} \,, \qquad C_{+-}  =  \epsilon^{uv}\partial_uX^- \partial_vX^a C_{+-a} \,,
\end{aligned}
\end{equation}
where $C_{+-a}=0$ is fixed by gauge invariance of the three-form and $C_{\pm ab}$ and $C_{abc}$ are assumed, in this work, to be nontrivial constants by background fixing. These constant components of the three-form have nontrivial contributions to the Hamiltonian when the embedding wrapps on a compactified sector of the target space, for example a flat torus. Let us also note that $X^-$ appears explicitly in the Hamiltonian through $C_a$ \cite{deWit}. Nevertheless, one may perform a canonical transformation of the Hamiltonian by performing the following change of variables \cite{mpgm6}
\begin{eqnarray}
   P_a=\widehat{P}_a-TC_a, \quad P_-=\widehat{P}_--TC_-, \quad S = \widehat{S}, \quad \widehat{X}^a=X^a, \quad \widehat{X}^-=X^-, \quad \widehat{\theta}=\theta. \nonumber 
\end{eqnarray}
Indeed, it can be seen it is a canonical transformation of the phase space variables, since it preserves the kinetic term and the Poisson brackets of the theory \cite{mpgm6} (notice that the notation that we are following here is different to the one used in \cite{mpgm6}). We may use the residual gauge symmetry generated by the constraints to impose the gauge fixing condition $P_-=P^0_-\sqrt{w}$, with $\sqrt{w}$ a regular density on the worldvolume. We can then eliminate ($X^-,P_-^0$) as canonical variables and obtain a formulation solely in terms of ($X^a,P_a$) and ($\theta,\bar{S}$).

If we consider a compactification of the target space, on a flat torus $T^2$ characterized by the Teichmuller parameter $\tau\in\mathbb{C}$ with $\mbox{Im}(\tau)> 0$ and a radius $R\in\mathbb{R}$, the embedding maps are splitted into the noncompact and compact sectors as follows $X^a=(X^m,X^r)$ with $m=1,\dots,7$ and $r=8,9$, respectively. We may perform a Hodge decomposition on the closed, but not exact, one-forms $dX^r=dX_h^r + dA^r$, where $dX_h^r$ are the harmonic one-forms and $dA^r$ are the exact ones. $dX_h^r$ may be written in terms of a normalized basis of harmonic one-forms $d\hat{X}^r$ as $dX_h^1+idX_h^2 = 2\pi R(l_r+m_r\tau)d\hat{X}^r$. The wrapping condition on the compact sector is given by
\begin{equation}
\label{ec2notas}
\oint_{\mathcal{C}_r} d \left(X^8 + iX^9 \right)= 2 \pi R \left(l_r + m_r \tau \right) \, \in  \mathcal{L} \,,
\end{equation}
where $\mathcal{C}_r$ denotes the homology basis on $\Sigma$, $\mathcal{L}$ is a lattice on the complex plane ($\mathbb{C}$) such that $T^2=\mathbb{C}/\mathcal{L}$ and the winding numbers $l_r,m_r$ defines the wrapping matrix
 \begin{eqnarray}
     \mathbb{W} = \left(\begin{array}{cc}
         l_8 & l_9 \\
          m_8 & m_9
     \end{array} \right). \label{windingmatrix}
 \end{eqnarray}
 Once the dependence on $X^-$ has been eliminated, a quantization condition on $C_{\pm}$ can be imposed. This condition corresponds to a 2-form flux condition on the target space 2-torus, whose pull-back through $X_h^r$, with $r=8,9$, generates a 2-form flux condition on the M2-brane worldvolume as follows \cite{mpgm10}
 \begin{eqnarray}\label{fluxpullback}
 \int_{T^2}C_{\pm}=\frac{1}{2} \int_{T^2}C_{\pm rs} d\widetilde{X}^r\wedge d\widetilde{X}^s = \frac{1}{2} \int_{T^2}C_{\pm rs} dX_h^r\wedge dX_h^s =c_{\pm}\int_\Sigma \widehat{F} = k_{\pm}A_{T^2},
 \end{eqnarray}
 where $C_{\pm rs}=c_{\pm}\epsilon_{rs}$ with $c_{\pm}\in \mathbb{Z}/\{0\}$, $\widetilde{X}^r$ are local coordinates on $T^2$, $k_{\pm}=nc_{\pm}$ with $n\in \mathbb{Z}/\{0\}$, $A_{T^2}=(2\pi R)^2\mbox{Im}(\tau)$ is the 2-torus area and  $\widehat{F}$ is a closed 2-form defined on $\Sigma$ such that it describes a worldvolume flux condition
 \begin{equation}\label{central charge}
   \int_{\Sigma}\widehat{F} = \frac{1}{2}\int_{\Sigma}dX^r \wedge dX^s \epsilon_{rs}=nA_{T^2},
  \end{equation}
  where the integer $n=det(\mathbb{W})\ne 0$ characterizing the irreducibility of the wrapping, where $\mathbb{W}$ is the winding matrix. Consequently, $C_{\pm}$ is a closed two-form defined on the target space torus. Indeed, the flux condition on $T^2$ implies a flux condition on $\Sigma$ which is known as 'central charge condition'. The irreducible wrapping condition ensures that the harmonic modes are nontrivial and independent.

The Hamiltonian of the M2-brane with $C_-$ fluxes becomes
\begin{eqnarray}
H^{C_-}&=&\frac{1}{2P^0_-}\int_\Sigma d^2\sigma \sqrt{w}\left[\Big(\frac{P_m}{\sqrt{w}}\Big)^2+\Big(\frac{P_r}{\sqrt{w}}\Big)^2 + \frac{T^2}{2}\left(\left\{X^m,X^n\right\}^2 + 2(\mathcal{D}_rX^m)^2 \right. \right. \nonumber \\
&+&\left. \left. (\mathcal{F}_{rs})^2+ (\widehat{F}_{rs})^2\right)\right]- \frac{T}{2P^0_-}\int_\Sigma d^2\sigma \sqrt{w} (\bar{\theta}\Gamma_-\Gamma_r\mathcal{D}_r\theta-T\bar{\theta}\Gamma_-\Gamma_m\left\{X^m,\theta\right\}),\nonumber \\ \label{HamiltonianM2}
\end{eqnarray}
which is equivalent to M2-brane with central charges \cite{Restuccia,mpgm6}, and 
\begin{eqnarray}\label{HamiltonianC+}
     H^{C_+} &=& H^{C_-} - 2\widehat{P}_-^0 T  \int d^2\sigma \sqrt{w}C_+,
\end{eqnarray}
only differs in a constant term \cite{mpgm6}.
Interestingly,  the supermembrane on $M_9\times T^2$ with a central charge condition associated with an irreducible wrapping is equivalent to the Hamiltonian of a supermembrane on $M_9^{LCG}\times T^2$ on a quantized $C_-$ background, i.e. $\mathcal{H}^{CC}=\mathcal{H}^{C_-}$.
The degrees of freedom of the theory are $X^m,A^r,\theta$. On the other hand, the symplectic covariant derivative is defined as \cite{Ovalle1}
\begin{eqnarray}
   \mathcal{D}_rX^m &=&D_rX^m+\left\{ A_r,X^m\right\},
\end{eqnarray}
with $D_r$ is a covariant derivative defined as \cite{mpgm2,mpgm7} and it satisfies 
\begin{eqnarray}
    (D_8+iD_9) \, \bullet  = 2\pi R (l_r+m_r\tau)\left\lbrace \widehat{X}^r,\, \bullet \right\rbrace , \nonumber 
\end{eqnarray}
where $\displaystyle \left\lbrace \bullet, \bullet \right\rbrace = \frac{\epsilon^{uv}}{\sqrt{w}}\partial_u \bullet \partial_v \bullet$. The gauge contribution is given by 
 $\widehat{F}$ the minimal curvature related to the flux on $\Sigma$ (\ref{fluxpullback}) and 
\begin{eqnarray}
  \mathcal{F}_{rs}&=& D_rA_s-D_sA_r+\left\{ A_r,A_s\right\}, \label{Fsymp}
\end{eqnarray}
corresponds to a symplectic curvature associated to the one-form connection $A_r dX^r$, where $A^r$ contains the dynamical degrees of freedom related to the exact sector of the map on $T^2$.

This Hamiltonian is subject to the local and global constraints associated to the area preserving diffeomorphisms (APD) 
\begin{eqnarray}
\small \left\{ \frac{P_m}{\sqrt{w}} , X^m\right\} + \mathcal{D}_r\left( \frac{P_r}{\sqrt{w}}\right)+\left\lbrace \frac{\bar{S}}{\sqrt{w}},\theta \right\rbrace  &\approx& 0, \label{LocalAPD}\\
 \oint_{C_S}\left[\frac{P_m dX^m}{\sqrt{w}} + \frac{P_r (dX_h^r+dA^r)}{\sqrt{w}} + \frac{\bar{S} d\theta}{\sqrt{w}}\right] &\approx& 0, \label{GlobalAPD}
\end{eqnarray}
which appears as a residual symmetry on the theory after imposing the LCG in the covariant formulation. In fact, we have shown that M2-branes with $C_{\pm}$ fluxes are invariant under the full group of simplectomorphisms, which considers the sectors connected and not connected to the identity. Furthermore, symplectomorphisms on $T^2$ are in one-to-one correspondence to symplectomorphisms on $\Sigma$  \cite{mpgm10}. Hence, the discreteness property of the latter automatically implies the discreteness of the M2-brane with $C_{\pm}$ fluxes. When $C_+\ne 0$, the spectrum is discrete and shifted by a constant value.

On \cite{mpgm12} a different canonical transformation of the phase space variables was considered on the M2-brane formulation, in order to eliminate the nonphysical degrees of freedom.  As a result, an equivalent M2-brane Hamiltonian with discrete supersymmetric spectrum was obtained with explicit presence of the transverse components of the three-form.

Classically, this Hamiltonian does not contain string-like spikes at zero cost energy that may produce instabilites  \cite{mpgm}. At quantum level the $SU(N)$ regularized theory has a purely discrete spectrum since it satisfy the sufficiency criteria for discreteness found in \cite{Boulton}. The theory preserves $1/2$ of the supersymmetry \cite{mpgm6}. This theory is equivalent or dual to the supermembrane with central charges. The M2-branes with $C_\pm$ fluxes can be formulated on twisted torus bundles with monodromy in $SL(2,Z)$ \cite{mpgm10}. In fact, the $U(1)$ principle bundle associated to the nontrivial quantized fluxes, or to the central charge condition, is compatible with the formulation of the M2-brane on a symplectic torus bundle, with structure group the symplectomorphism which preserve the $U(1)$ curvature. There exists a natural homomorphism
\begin{eqnarray}\label{monodromy}
    \mathcal{M}_G:\Pi_1(\Sigma)\rightarrow \Pi_0(Symp(T^2))=SL(2,Z).
\end{eqnarray}
The subgroup of $SL(2,Z)$ determined by the homomorphism is called the monodromy of the formulation. The classification of symplectic torus bundles with monodromy in terms of $H^2(\Sigma, \mathbb{Z}^2_{\rho})$ was found by \cite{Kahn}. In the aforementioned paper it is shown the existence of a one-to-one correspondence between the inequivalent classes of symplectic torus bundles for a given monodromy conjugacy class inducing the module structure $Z_\rho^2$ on $H_1(T^2)$ and the elements of $H^2(\Sigma,Z_\rho^2)$, the second cohomology group of the bundle with base $\Sigma$ and coefficients in $Z_\rho^2$. This homomorphism gives to each homology and coholomogy group on the bundle the structure of $Z\left[\Pi_1(\Sigma)\right]$-module.  It classifies the symplectic torus bundles for a given monodromy in terms of the characteristic class. Hence, the symplectic torus bundles, with base manifold a torus, are
classified, for a given monodromy, according to the inequivalent coinvariants \cite{mpgm7,mpgm2}.

Therefore, sectors of M2-branes on $M_9\times T^2$ with the irreducible wrapping condition, contain two compatible gauge structures. The first one is given by the symplectic structure of the bundle, which ensures the existence of a symplectic connection under symplectomorphisms. The second gauge structure is a nontrivial $U(1)$ principal bundle related to the 2-form flux on $\Sigma$ due to the central charge condition or the 2-form flux condition on the target-space.

In \cite{mpgm10} it was proved that the symplectic structure and the U(1) principal bundle are related and generate a twisted torus bundle,
\begin{equation}
\label{ec7notasseccion6.2}
\mathbb{T}_W^3\equiv {T}_{U(1)}^2 \rightarrow E' \rightarrow \Sigma \, ,  
\end{equation}
where the base manifold is given by the worldvolume Riemann surface $\Sigma$, the fiber is a twisted torus $\mathbb{T}^3$, given by the U(1) principal bundle associated with the nontrivial flux condition on $T^2$.

The LCG Hamiltonian of an M2-brane with $C_{\pm}$ fluxes, can be generalized to make  the presence of the supergravity three-form transverse components, $C_{abc}$ with $a=(m,r)$,  explicit in the final Hamiltonian \cite{mpgm14}. This is relevant to make manifest in its D-brane description, subject to quantized RR and NSNS forms, the appearance of the transverse components of the B-field in the associated DBI terms.  However, both nontrivial sectors can be shown to be equivalent due to canonical transformations \cite{mpgm13}.	
	
	
	\section{$SL(2,Z)$  ($p,q$)-strings from the M2-brane with $C_{\pm}$ fluxes.}\label{sec3}

In this section, we extend to the supersymmetric M2-brane with nontrivial $C_\pm$ fluxes and trivial monodromy, \cite{Schwarz6} and \cite{mpgm21}. See \cite{Uehara} for a different approach. In Section 4, we will consider the case with nontrivial monodromy. We will show, in this section, that the mass operator of type IIB $SL(2,Z)$ ($p,q$)-strings compactified on a circle of radius $R_B$ coincides with the mass operator of the M2-branes on a $T^2$ with central charges, or equivalently with $C_-$ fluxes. The irreducible wrapping condition that characterizes these sectors, ensures the existence of bound states. The sector without the central charge condition is only able to reproduce type IIB fundamental strings, $(1,0)$-strings on $M_9\times S^1$ with null Kaluza Klein on the compact sector, but with non zero winding. We will show that the results found in \cite{Schwarz6} are only valid when the central charge or equivalently the $C_-$ flux condition is present. We extend those results to include the supersymmetric sector and the Hamiltonian terms of the M2-brane to reproduce  the $(p,q)$-string mass operator. A detailed computation will be performed to facilitate the understanding of the differences with the new $(p,q)$ string sector discussed in section 5.

\paragraph{$SL(2,Z)$ symmetries on the supermembrane with $C_{\pm}$ fluxes.}

In \cite{mpgm3} two inequivalent $SL(2,Z)$ symmetries of the M2-brane with central charges were identified. One is associated with the target torus and will be denoted as $SL(2,Z)_{T^2}$, while the other is associated with the base manifold and will be denoted as $SL(2,Z)_{\Sigma}$. In \cite{mpgm10} M2-branes  with $C_{\pm}$ fluxes were shown to be  invariant under the 2-dimensional area preserving diffeomorphisms, or equivalently, 2-dimensional symplectomorphisms, connected and not connected to the identity. The invariance of the Hamiltonian under those connected with the identity is guaranteed by the first class constraint of the theory. The isotopy classes of symplectomorphisms on the base manifold determine a group, which in this case is $SL(2,Z)_{\Sigma}$. The symplectomorphisms not connected to the identity change the homology basis on $\Sigma$ together with the corresponding basis of harmonic one-forms and the winding matrix as follows
	\begin{eqnarray}
	    d\widetilde{X}^r \rightarrow (S_1^*)^r_sd\widehat{X}^s \label{SL(2,Z)sigma1},\quad
	    \mathbb{W} \rightarrow \mathbb{W}(S_1^*)^{-1} , \label{SL(2,Z)sigma2}
	\end{eqnarray}
	with $S_1^*\in SL(2,Z)_\Sigma$.
The symplectomorphisms not connected with the identity on target $T^2$ are the ones that change the moduli of the 2-torus by a modular transformation \cite{mpgm3} as follows,
\begin{eqnarray}
	\tau&\rightarrow& \tau'=\frac{a\tau+b}{c\tau+d} , \quad
	R\rightarrow R'= R|c\tau+d| , \quad
	A\rightarrow A'= Ae^{i\varphi_\tau}, \nonumber \\	
	\mathbb{W}&\rightarrow&\mathbb{W}'=S_2^*\mathbb{W},  \quad
	Q\rightarrow Q'=S_2Q,	\label{SdualQ}
	\end{eqnarray}
with $S_2$,$S_2^*$ matrices of $SL(2,Z)_{T^2}$ given by
	\begin{eqnarray}
	S_2 = \left(\begin{array}{cc}
	  a   & b \\
	    c & d
	\end{array}\right), \quad
	S_2^* = \left(\begin{array}{cc}
	  a   & -b \\
	    -c & d
	\end{array}\right), \quad
	    c\tau+d = |c\tau+d|e^{-i\varphi_\tau} . \label{Sdualphi} 
	\end{eqnarray}
The bosonic part of the Hamiltonian is invariant under (\ref{SdualQ}), and it  corresponds to the action of $S_U$-duality on M2-branes on a torus \cite{mpgm3}.

 The full supersymmetric Hamiltonian also becomes invariant under (\ref{SdualQ}) if the following transformation is added
\begin{eqnarray}
   \Gamma &\rightarrow& \Gamma'=\Gamma e^{i\varphi_\tau},
\end{eqnarray}
 where $\Gamma=\Gamma_8+i\Gamma_9$ is the complex gamma matrix present in the fermionic term and related to the compact directions.
 
While the previous $SL(2,Z)_{T^2}$ is generic for a M2-brane on a 2-torus, the $SL(2,Z)_\Sigma$ transformation is characteristic of sectors of M2-brane generated by the $C_\pm$ flux condition. Therefore, the M2-brane with $C_{\pm}$ fluxes Hamiltonian is invariant under both $SL(2,Z)$ transformations. It is worth to mention that both transformations are independent. The irreducible wrapping condition ensures a one-to-one correspondence of symplectomorphisms on $T^2$ and $\Sigma$, also assumed to be a 2-torus.
\subsection{Mass operator of the supermembrane with $C_{\pm}$ fluxes}
We will firstly consider the winding and KK sectors of the mass operators \cite{mpgm21}. The embedding map to the compact sector is defined as  
 \begin{eqnarray}\label{generalmap}
 dX=(2\pi R)(l_s+m_s\tau)d\widehat{X}^s+dA. 
 \end{eqnarray}
 However, as noticed in \cite{mpgm21}, it is possible to use the independent and arbitrary $SL(2,Z)$ symmetries on $T^2$ and $\Sigma$ to rewrite the winding matrix (\ref{windingmatrix}) as
 \begin{eqnarray}\label{Simplifiedwinding}
     \mathbb{W} &=& \left( \begin{array}{cc}
         n & 0 \\
         0 & 1
     \end{array}\right),
 \end{eqnarray}
and therefore (\ref{generalmap}) becomes $dX=2\pi R (n d\widehat{X}^{1}+ \tau d\widehat{X}^{2})+dA$, where $dA=dA^1+idA^2$ is a dynamical exact one-form.

 The pure harmonic contribution associated with the wrapping on the  M2-brane Hamiltonian is given by
 \begin{eqnarray}
    \frac{T^2}{P_-^0} \int d^2\sigma \left[\frac{1}{4}\sqrt{w}\left\lbrace X_h^r,X_h^s \right\rbrace^2 \right] = \frac{1}{2P_-^0}(TnA_{T^2})^2,
 \end{eqnarray}

  with $n=det(\mathbb{W})$. Therefore, the winding term on the mass operator of the M2-brane with $C_-$ fluxes is
  \begin{eqnarray} \label{massoperatorwinding}
    M_{C_\pm}^2 = (TnA_{T^2})^2 + \dots,
 \end{eqnarray}
 However, the mass operator of the M2-brane with $C_+$ fluxes contains an extra harmonic contribution that results in a constant term
 \begin{eqnarray} \label{massoperatorwinding2}
    M_{C_+}^2 = (TnA_{T^2})^2- 2P_-^0 Tk_+A_{T^2} + \dots,
 \end{eqnarray}
 with $k_+=nc_+$. As the irreducible wrapping condition guarantee that $n\neq 0$, then the winding term is strictly related with these sectors. A reducible wrapping will not generate those terms. 
 
 In order to reproduce the KK term on the mass operator, the zero modes of the momentum in the compact sector can be expressed, following \cite{mpgm21}, as 
 \begin{eqnarray}
P^0_{r}=\int_\Sigma p_r d\sigma^1\wedge d\sigma^2,
\end{eqnarray}
with $r=8,9$, which can be rewritten in terms of the Hodge dual of two well-defined associated 2-forms $(F)^r$ on $\Sigma$. In fact, fixing $r$, it can be seen that
for each r, \begin{eqnarray}
Rp&=& b\sqrt{w}\left(\star F\right),
\end{eqnarray}
with $\displaystyle \star F = \frac{\epsilon^{uv}F_{uv}}{2\sqrt{w}}$ and $b$ a proportionality constant with dimensions of $(energy)\times (length)$. If $c=\hbar=1$, then $b=1$. Consequently
\begin{eqnarray}
RP^0 &=& b\int_\Sigma F,
\end{eqnarray}
and then, the following quantization conditions, are imposed for each value of $r$
\begin{eqnarray}
\int_\Sigma F_r = \widehat{m}_r \in \mathbb{Z}.
\end{eqnarray}
In order to guarantee that the maps from the base manifold to the compact target sector are from circles onto circles, we must consider the left hand member of
\begin{eqnarray}\label{circletocirlce}
\frac{1}{2\pi R}\oint_{C_S}\mathbb{M}^{-1}\left(\begin{array}{c}
    dX^8 \\
    dX^9
\end{array}\right) = \oint _{C_S}\mathbb{W}\left(\begin{array}{c}
    d\widehat{X}^8 \\
    d\widehat{X}^9
\end{array}\right),
\end{eqnarray}
with
\begin{eqnarray}\label{matrixM}
\mathbb{M} = \left( \begin{array}{cc}
    1 & \mbox{Re}(\tau) \\
     0 & \mbox{Im}(\tau)
\end{array} \right),
\end{eqnarray}
where $\mathbb{W}$ given by (\ref{Simplifiedwinding}),  satisfies $det (\mathbb{W})=n\ne 0$. Now, by using the corresponding conjugate momenta, 
\begin{eqnarray}\label{Conju_Momenta}
RP^0_s\mathbb{M}^s_r=R\int_\Sigma p_s\mathbb{M}^s_rd\sigma^1\wedge d\sigma^2 = b\widehat{m}_r.
\end{eqnarray}
Consequently, the KK modes are given by
$
\displaystyle P^0_r = b(\mathbb{M}^{-1})^s_{r}\frac{\widehat{m}_s}{R}.
$
Hence,
\begin{eqnarray}
P^0_8 = b\frac{\widehat{m}_8}{R},\label{P01} \quad P^0_9 = b\frac{\widehat{m}_9-\widehat{m}_8\mbox{Re}(\tau)}{R\mbox{Im}(\tau)}\, .\label{P02}
\end{eqnarray}
 The KK contribution to the mass operator is given by
 \begin{eqnarray}\label{KKterma}
2P^0_- \left( \frac{1}{2P^0_-}P^0_rP^{0r}\right) = b^2 m^2\frac{\vert q\tau-p \vert^2 }{(R\mbox{Im}(\tau))^2} ,
\end{eqnarray}
with $q$, $p$ relatively primes, where it has been used that   $\widehat{m}_8=mq$ and $\widehat{m}_8=mp$, with $m\in\mathbb{Z}$. In fact, it can be checked that the KK term is $SL(2,Z)_{T^2}$ invariant if $p,q$ transform according to (\ref{SdualQ}).  The expression of the KK-term given by (\ref{KKterma}), which is in agreement with the one obtained in \cite{Schwarz},  is strictly related to a well-defined compactification on a 2-torus, i.e. it is associated to the irreducible wrapping condition present in well-behaved sectors of M2-branes. Indeed, when the wrapping of the M2-brane on the compact sector is reducible, $det(\mathbb{W})=0$, the map from $\Sigma$ to $T^2$ becomes degenerate and there is no holomorphic map between them, consequently, there is no map from circles to circles. 

 Finally, the M2-brane with $C_\pm$ fluxes mass operator corresponds to \cite{mpgm22} 
 \begin{eqnarray}\label{MassOp_MonTriv}
    \mathcal{M}_{C_\pm}^2 &=& (TnA_{T^2})^2 + b^2 \frac{m^2\vert q\tau-p \vert^2 }{(R\mbox{Im}(\tau))^2}+2\widehat{P}_-^0H'^{C_\pm},
 \end{eqnarray}
 where
\begin{eqnarray}
H'^{C_-}&=&\frac{1}{2P^0_-}\int_\Sigma d^2\sigma \sqrt{w}\left[\Big(\frac{P'_m}{\sqrt{w}}\Big)^2+\Big(\frac{P'_r}{\sqrt{w}}\Big)^2 + \frac{T^2}{2}\left(\left\{X^m,X^n\right\}^2 + 2(\mathcal{D}_rX^m)^2 \right. \right. \nonumber \\
&+&\left. \left. (\mathcal{F}_{rs})^2\right)\right]- \frac{T}{2P^0_-}\int_\Sigma d^2\sigma \sqrt{w} (\bar{\theta}\Gamma_-\Gamma_r\mathcal{D}_r\theta-T\bar{\theta}\Gamma_-\Gamma_m\left\{X^m,\theta\right\}),  \label{Hamiltonian_prime_Cmenos}\\
H'^{C_+} &=& H'^{C_-} - 2\widehat{P}_-^0 TnA_{T^2}c_{+}, \label{Hamiltonian_prime_Cmas}
\end{eqnarray}
The prime on the fields in the Hamiltonian indicates that the zero modes and the pure harmonic contributions are excluded. The winding and the general expression for the KK contribution on this Hamiltonian were obtained in \cite{Schwarz6}. They are strictly related to the M2-brane with a central charge condition associated with the irreducibility of the wrapping or with  the presence $C_\pm$ fluxes,  on $M_9\times T^2$. The irreducible wrapping condition, ensures the appearance of both terms.  

\subsection{Mass operator for the type $IIB$ ($p,q$)-string}

Now we will show that the full mass operator of the $(p,q)$ string, is directly related to the irreducible wrapping condition induced by the $C_{\pm}$ flux, that determines its characteristic tension $T_{(p,q)}$. In order to reproduce the stringlike excitations on the M2-brane mass operator we assume the dynamical variables to depend  only on a linear combination of the two spatial coordinates. Instead of considering the local coordinates ($\sigma^1,\sigma^2$), we will work  with the minimal maps $\widehat{X}^r$ given by (\ref{circletocirlce}), that is $
\sigma^1,\sigma^2 \rightarrow \widehat{X}^8,\widehat{X}^9
$. The Jacobian of the transformation is given by $det(J(\sigma^1,\sigma^2))=\sqrt{w}$ where
\begin{eqnarray}
   \sqrt{w}=\frac{1}{2}\epsilon^{uv}\partial_u \widehat{X}^r
\partial_v \widehat{X}^s\epsilon_{rs},
\end{eqnarray}
is nonsingular over $\Sigma$. Therefore
$
\sqrt{w}d\sigma^1\wedge d\sigma^2 = d\widehat{X}^8\wedge d\widehat{X}^9
$. Let us now define string configurations  such that $\Phi (c\tau,\sigma^1,\sigma^2)=\Phi (c\tau,\rho)$ with $\Phi=(X^m,A^r,\theta)$ fields of the theory and $\rho = q_1\widehat{X}^8 + q_2\widehat{X}^9$ being $q_1$, $q_2$ relatively primes. In that case, we have that
\begin{eqnarray}
\{X^{m},X^{n}\}=\{X^{m}, A^{r}\}=\{A^{r},A^{s}\}=\{ X^m,\theta\}=\{A^r,\theta\}= 0,
\end{eqnarray}
and the Hamiltonian $H'^{C_-}$ (\ref{Hamiltonian_prime_Cmenos}), on the string configurations, can be written as
\begin{eqnarray}
 H'_{C_-}\vert_{SC} &=& \frac{1}{2P_-^0}\int d^2\sigma \sqrt{w} \left\lbrace \frac{(P'_m)^2}{w} +\frac{(P'_r)^2}{w} + T^2\left\lbrace X_h^r,X^m \right\rbrace^2   \right. \nonumber \\
&+& \left. T^2\left\lbrace X_h^r,A^s \right\rbrace^2+2P^0_- T\Bar{\theta}\Gamma^-\Gamma_r\left\lbrace X_h^r,\theta \right\rbrace \right\rbrace   .
\end{eqnarray}
This Hamiltonian is subject to the usual local (\ref{LocalAPD}) and global (\ref{GlobalAPD}) constraints on the string configurations,
\begin{eqnarray}
\left\lbrace \frac{P'_r}{\sqrt{w}},X_h^r\right\rbrace &\approx& 0, \label{LocalconstraintSC} \\
\oint_{C_S}\left[\frac{P_m dX^m}{\sqrt{w}} + \frac{P_r (dX_h^r+dA^r)}{\sqrt{w}} + \frac{\bar{S} d\theta}{\sqrt{w}}\right] &\approx& 0, \label{globalconstraintSC}
\end{eqnarray}
where $C_s$ is the homology basis dual to the harmonic maps $\widehat{X}^r$. It can be checked using the Jacobi identity, that the local constraint (\ref{LocalconstraintSC}) can be solved to obtain
\begin{eqnarray}\label{PrPI}
\frac{P'_r}{\sqrt{w}} = T\epsilon_{rs}\left\lbrace X_h^s,\frac{\Pi}{\sqrt{w}} \right\rbrace.
\end{eqnarray}
In terms of a new pair of canonical variables ($X^*,P_*$)
\begin{eqnarray}
X^* = \frac{\Pi}{\sqrt{w}}, \label{newvariable1} \quad
P_* = T\sqrt{w}\left\lbrace X_h^r,A^s\right\rbrace \epsilon_{rs}, \label{newvariable2}
\end{eqnarray}
 the kinetic terms associated with the compact sector become
\begin{eqnarray}
\frac{1}{2}\left( \frac{P'_r}{\sqrt{w}}\right)^2 = \frac{T^2}{2}\left\lbrace X_h^r,X^*\right\rbrace^2 , \quad
\frac{T^2}{2}\left\lbrace X_h^r,A^s\right\rbrace^2 = \frac{1}{2}\left( \frac{P_*}{\sqrt{w}}\right)^2,
\end{eqnarray}
 Consequently, we have that
\begin{eqnarray}
H_{C_-}\vert_{SC} &=& \frac{1}{2P_-^0}\int d\widehat{X}^8 \wedge d\widehat{X}^9 \left\lbrace \left(\frac{P'_M}{\sqrt{w}}\right)^2 + T^2\left\lbrace X_h^r,X^M \right\rbrace^2  \right. \nonumber \\
&+& \left.   2P^0_- T\sqrt{w}\Bar{\theta}\Gamma^-\Gamma_r\left\lbrace X_h^r,\theta \right\rbrace \right\rbrace. \label{hamiltonianSC}
\end{eqnarray}
where $X^M=(X^m,X^*)$ and $M=1,\dots,8$. The total time derivatives have been eliminated from the Hamiltonian formulation. This expression corresponds to a susy harmonic oscillator \cite{Duff6}. The bosonic and fermionic potentials can be expressed in complex notation as
\begin{eqnarray}
\frac{1}{2}\left\lbrace X_h^r,X^M \right\rbrace^2 &=& \frac{1}{2}\vert \left\lbrace X_h , X^M\right\rbrace \vert^2, \label{Bosonicpotential} \\
\Bar{\theta}\Gamma^-\Gamma_r\left\lbrace X_h^r,\theta \right\rbrace &=& \frac{1}{2}\Bar{\theta}\Gamma^-\left[ \Bar{\Gamma}\left\lbrace X_h,\theta \right\rbrace + \Gamma\left\lbrace \Bar{X}_h,\theta \right\rbrace \right]. \label{fermionicpotential}
\end{eqnarray}
In order to express as a string theory Hamiltonian, let us perform a change on the canonical basis of homology, with its corresponding change on the  harmonic on the basis of harmonic one-forms,
\begin{eqnarray}
d\widetilde{X}^8 = q_1 d\widehat{X}^8 + q_2 d\widehat{X}^9 , \label{homologybasis1} \quad
d\widetilde{X}^9 = nq_3 d\widehat{X}^8 + q_4 d\widehat{X}^9. \label{homologybasis2}
\end{eqnarray}
with $q_1$ prime relative to $q_2$ and $n$. It can be seen that there always exist $q_3$ and $q_4$  such that
\begin{eqnarray}
\left( \begin{array}{cc}
    q_1 &  q_2 \\
     nq_3 & q_4
\end{array}\right) \in SL(2,Z),
\end{eqnarray}
with
$
d\widetilde{X}^8\wedge d\widetilde{X}^9 = d\widehat{X}^8\wedge d\widetilde{X}^9
$. We can use the $SL(2,Z)_{T^2}$ and $SL(2,Z)_{\Sigma}$ to rewrite the Hamiltonian in such a way that the winding matrix (\ref{Simplifiedwinding}) remains invariant. Therefore, the modulus of the harmonic 1-form remains invariant if \footnote{It can be seen that the transformation of the complex harmonic one-form of (\ref{generalmap}) is given by
\begin{eqnarray}\label{Harmonic_transformation}
    dX_h = (2\pi R)(nd\widehat{X}^8+\tau d\widehat{X}^9) = (2\pi R')(nd\widetilde{X}^8+\tau' d\widetilde{X}^9) e^{- i\varphi}=d\widetilde{X}_he^{- i\varphi},
\end{eqnarray}
with 
$\displaystyle  e^{i\varphi}= \frac{q_4-q_3\tau}{\vert q_4-q_3\tau \vert}
$.}
\begin{eqnarray}
R' = R\vert q_4-q_3\tau \vert, \quad
\tau' = \frac{q_1\tau-nq_2}{q_4-q_3\tau},
\end{eqnarray}
The Hamiltonian written in the new variables becomes, 
\begin{eqnarray}
H'_{C_-}\vert_{SC} &=& \frac{1}{2P_-^0}\int d\widetilde{X}^8\wedge d\widetilde{X}^9 \left\lbrace \left(\frac{P'_M}{\sqrt{w}}\right)^2 + T^2(2\pi R'\vert \tau' \vert)^2\partial_8X^M \partial^8X_M  \right.  \nonumber \\
&-& \left. 2P^0_- T(2\pi R')\Bar{\theta}\Gamma^-\left[ (\widetilde{\Gamma}_8 \mbox{Re}(\tau')+\widetilde{\Gamma}_9 \mbox{Im}(\tau')) \right]\partial_8\theta \right\rbrace, \label{H_Before_Farkas}
\end{eqnarray}
with
\begin{eqnarray}
    \widetilde{\Gamma}_8 &=& \frac{1}{\vert q_4-q_3\tau \vert} \left[ \Gamma_8 (q_4-q_3 \mbox{Re}(\tau))-\Gamma_9q_3\mbox{Im}(\tau) \right] , \\
    \widetilde{\Gamma}_9 &=& \frac{1}{\vert q_4-q_3\tau \vert} \left[ \Gamma_8q_3\mbox{Im}(\tau) +\Gamma_9 (q_4-q_3 \mbox{Re}(\tau))\right],
\end{eqnarray}
such that 
\begin{eqnarray}
    (\widetilde{\Gamma}_8)^2 &=& (\widetilde{\Gamma}_9)^2 = \mathbb{I}, \\
    \left\lbrace \widetilde{\Gamma}_8 , \Gamma_m \right\rbrace &=& \left\lbrace \widetilde{\Gamma}_9 , \Gamma_m \right\rbrace = 
    \left\lbrace \widetilde{\Gamma}_8 , \widetilde{\Gamma}_9 \right\rbrace = 0.
\end{eqnarray}
where $\left\lbrace , \right\rbrace$ denotes the anticommutator. Using the proposition III.2.3 from \cite{Farkas}, it can be seen that we can rewrite the Hamiltonian as
\begin{eqnarray}
H'_{C_-}\vert_{SC} &=& \frac{1}{2P_-^0}\int d\widetilde{X}^8 \left\lbrace \left(\frac{P'_M}{\sqrt{w}}\right)^2 + T^2(2\pi R'\vert \tau' \vert)^2\partial_8X^M \partial^8X_M  \right.  \nonumber \\
&-& \left. 2P^0_- T(2\pi R')\Bar{\theta}\Gamma^-\left[(\widetilde{\Gamma}_8 \mbox{Re}(\tau')+\widetilde{\Gamma}_9 \mbox{Im}(\tau')) \right]\partial_8\theta \right\rbrace.
\end{eqnarray}
Finally, the global constraint remains to be solved (\ref{globalconstraintSC}). It can be checked that the constraint related to $\widetilde{X}^9$ leads  to $\widehat{m}'_9=0$, where the prime indicates the transformation under $SL(2,Z)_{T^2}$. Therefore, $q=1$ and $p=0$. In order to verify that the global constraint due to $\widetilde{X}^8$ reproduce the level matching condition, let us recall that $\widetilde{X}^8$ is adimensional. Therefore, we define
\begin{eqnarray}\label{sigmadimensions}
\xi = a\widetilde{X}^8 + C,
\end{eqnarray}
where 
$\label{Adimensionslength}
a= \frac{K\sqrt{P_-^0}}{\widetilde{T}},
$
with $\widetilde{T}= T(2\pi R')\vert \tau' \vert$. On this expression $K$ is a constant with dimensions of $(energy)^{1/2}$, the constant $C$ does not depend on $\widetilde{X}^8$ and $a$ has dimensions of $(length)$. Then,
$
d\widetilde{X}^8 = a^{-1}d\xi, \quad
\partial_8 = a\partial_\xi,
$
and by demanding the kinetic term to remain invariant the Hamiltonian becomes
\begin{equation}
 H'_{C_-}\vert_{SC} = \frac{K}{\sqrt{P_-^0}}\int d\xi \frac{1}{2} \left\lbrace \frac{1}{\widetilde{T}}\left(\frac{P_M^*}{\sqrt{w}}\right)^2 +  \widetilde{T} \partial_\xi X^M \partial_\xi X_M - \frac{\bar{S}^*}{\sqrt{w}}\Gamma^*\partial_\xi\theta \right\rbrace, \label{Happendix1}
\end{equation}
with 
\begin{eqnarray}
   \Gamma^* &=& \frac{\widetilde{\Gamma}_8 \mbox{Re}(\tau')+\widetilde{\Gamma}_9 \mbox{Im}(\tau')}{\vert \tau'\vert}.
\end{eqnarray}
It can be seen that $(\Gamma^*)^2 = \mathbb{I}$ and $\left\lbrace \Gamma^* , \Gamma^m \right\rbrace = 0$,  therefore $(\Gamma^+,\Gamma^-,\Gamma^M)$ with $\Gamma^M=(\Gamma^m,\Gamma^*)$ satisfy the Clifford algebra. The Hamiltonian of the string configurations is invariant under the supersymmetry transformations inherited from M2-brane theory in the LCG.

By using the $SO(8)$ spinor decomposition shown in the appendix, the fermionic contribution can be re-expressed as
\begin{eqnarray}
\frac{\bar{S}^*}{\sqrt{w}}\Gamma^*\partial_\xi\theta &=& i \sqrt{2} P_-^0 
\left[ \chi^+ \partial_\xi \chi^- + \chi^-\partial_\xi \chi^+\right],
\end{eqnarray}
where $\chi^\pm$ are spinors whose 8 components are given by the nontrivial components of the $SO(9)$ spinors $\psi^\pm$. Using $SO(7)$ spinors 
$\lambda^1 = \chi^++\chi^-, \quad
    \lambda^2 = \chi^+-\chi^-,$  and re-scaling them as
 \begin{eqnarray}
   \lambda^1 \rightarrow \widehat{\lambda}^1 = 2^{1/4}\sqrt{\widehat{P}_-^0}\lambda^1, \quad
   \lambda^2 \rightarrow \widehat{\lambda}^2 = 2^{1/4}\sqrt{\widehat{P}_-^0}\lambda^2, \label{SO(7)reescaled}
\end{eqnarray}   
we obtain that the Hamiltonian can now be written as
\begin{equation}\label{Hamiltoniantrivialmonodromy1}
\small H_{C_-}\vert_{SC} = \frac{1}{2} \frac{K}{\sqrt{P_-^0}}\int d\xi \left\lbrace \frac{1}{\widetilde{T}}\left(\frac{P_M^*}{\sqrt{w}}\right)^2 +  \widetilde{T} \partial_\xi X^M \partial_\xi X_M-  \frac{i}{a} (\widehat{\lambda}^1\partial_\xi \widehat{\lambda}^1 - \widehat{\lambda}^2\partial_\xi \widehat{\lambda}^2)  \right\rbrace.
\end{equation}
 The mass operator of the stringlike configurations associated with $C_{\pm}$ fluxes can now be written as
	\begin{equation}
	\begin{aligned}
    M^2_{C_-}\vert_{SC} &=&  (TnA_{T^2})^2 + \frac{m^2\vert \tau'\vert^2 }{(R'\mbox{Im}(\tau'))^2} +  T8\pi^2 R'\vert \tau'\vert(N_T+\bar{N}_T), \label{MassOperatorFinal}
    \end{aligned}
\end{equation}
and 
\begin{equation}
	\begin{aligned}
    M^2_{C_+}\vert_{SC} &=& M^2_{C_-}\vert_{SC} -2\widehat{P}_-^0 TA_{T^2}k_{+}, \label{MassOperatorFinalC_mas}
    \end{aligned}
\end{equation}
where $c=\hbar=1$, and $N_T$, $\bar{N}_T$ are the total number operators defined on the appendix B.

Since $\tau'$ denotes an arbitrary point on the upper complex plane, in order write the Hamiltonian in terms of the fundamental domain,  a modular transformation is performed with
\begin{eqnarray}
    \left( \begin{array}{cc}
        q & -p \\
        Q & P
    \end{array} \right) \in SL(2,Z),
\end{eqnarray}
where the minus sign is convention and $q$, $p$ are relatively primes. The mass operator of the stringlike configurations associated with the M2-brane with $C_{\pm}$ fluxes can be written as
	\begin{equation}
    M_{C_-}^2\vert_{SC} =  (T_{11}nA_{T^2})^2 + \left( \frac{m\vert q\tau-p\vert}{R \mbox{Im}(\tau)}\right)^2 + T8\pi^2 R\vert q\tau-p\vert(N_T+\bar{N}_T) \label{MassOperatorFinal2}
\end{equation}
and as $A_{T^2}$ remains invariant, we have that $M^2_{C_+}\vert_{SC}$ is given by (\ref{MassOperatorFinalC_mas}). 
 We must emphasize that the winding and KK term on (\ref{MassOperatorFinal}) are strictly related to sectors of M2-brane in the light-cone gauge on $M_9\times T^2$ with consistent quantum behaviour. When the irreducible wrapping condition is not satisfied, we have shown that the winding and KK term are not reproduced.

The last term on (\ref{MassOperatorFinalC_mas}) is a constant contribution due to the flux condition on $C_+$. This term does not appear in the M2-brane with central charges, or equivalently in the M2-brane with $C_-$ fluxes. In all cases the mass operator (\ref{MassOperatorFinalC_mas}) is invariant under the full $SL(2,Z)_{T^2}$ symmetry.

Using (\ref{newvariable1}) and then the expansions (\ref{BosonicBC88}) and (\ref{FermionicBC}) on the appendix B, with $c=\hbar=1$ it is possible to obtain from the global constraint the level matching constraint as
\begin{eqnarray}\label{LMC}
\bar{N}_T-N_T &=& \widehat{m}_8 n,
\end{eqnarray}
Recalling that the type $IIB$ mass operator ($p,q$)-string compactified on a $S^1$ of radius $R_B$, is obtained from the supermembrane by using $M^2=\beta^2M_{{(p,q)}}^2$ \cite{Schwarz6}, with 
\begin{eqnarray}
\tau=\lambda_0,\quad \beta^2 = \frac{TA_{T^2}^{1/2}}{T_c}, \quad
 R_B^2 = (TA_{T^2}^{3/2}T_c)^{-1}, \label{tau,beta_y_R}
 \end{eqnarray}
 where $T_c = T^{2/3}$ is the string tension.
Taking into account that the wrapping terms on the 11D formulation side correspond to the KK contribution on the type IIB side and vicecersa, assuming the $C_+=0$ flux contribution, by substitution, one recovers the compactified type IIB mass operator ($p,q$)-string
\begin{eqnarray}\label{cuerda IIB}
 M^2_{(p,q)} = \left( \frac{n}{R_B} \right)^2 + (2\pi R_B m T_{(p,q)})^2 + 4\pi T_{(p,q)}(N_L + N_R),
\end{eqnarray}
where the tension of the ($p,q$)-string is 
\begin{eqnarray}
 T_{(p,q)} = \frac{\vert q\lambda_0-p\vert}{(\mbox{Im}(\lambda_0))^{1/2}}T_c, \label{Tpq}
\end{eqnarray}
with $T_c$ the tension of the string, $\lambda = \xi + i\exp{\phi}$  the axion-dilaton of the type IIB theory with $\phi$ correspond the dilaton field and $\lambda_0$ is the scalar corresponding to the asymptotic value of $\lambda$.

For the general case with $C_+\ne 0$  the only difference with (\ref{cuerda IIB})  will be a constant shift on the ($p,q$)-string mass operator given by  $2P_-^0T_c^{1/6}R_B^{-2/3}k_+$.

It can be seen that the central charge condition is directly necessary to obtain the KK contribution but also allowing to define the $T_{(p,q)}$ for $p,q\ne 0$ since it requires a proper map on a torus.
Bound states of ($p,q$)-strings are strictly related to sectors of M2-brane on $M_9\times T^2$ with irreducible wrapping. The sector with $n=0$ is only able to reproduce wrapped type IIB ($1,0$)-strings with null KK contribution. The low energy limit is given by maximal supergravity in 9D for any value of $n$.



\section{Mass operator of the M2-brane with monodromy}

In this section, we obtain the supersymmetric mass operator corresponding to the M2-branes with fluxes and nontrivial monodromy. We will consider only parabolic monodromies, and analysis for other monodromies will be presented elsewhere. Let us emphasize that the contribution of the monodromy is nontrivial. The mass operator in \cite{Schwarz6} was obtained for M2-branes on a torus with trivial monodromy, in this case each pair of charges ($p,q$) determines a conivariant. These coinvariants are related among them by the $SL(2,Z)$ symmetry. The main point is that when when the monodromy is not trivial, the coinvariants contain a set of ($p,q$) charges related by a internal symmetry that leaves the coinvariant invariant. Furtheremore, there is also a symmetry relating the different coinvariants among them. In what follows we will determine both symmetries, the formulation reduces to the one in previous sections when the monodromy is trivial. Let us notice that for parabolic monodromies, the coinvariants are defined solely by the charge $q$ from the pair ($p,q$). That is, different $q$ define different coinvariants. All pairs ($p,q$) with the same $q$ belong to the same coinvariant. The mass operator depends also on the moduli and winding number. In particular we have to give the associated moduli to the corresponding coinvariant. In the trivial monodromy case one has to provide the moduli for each pair ($p,q$). In the parabolic monodromy case, given $q$ the internal symmetry define an equivalence class of charges $p$ and moduli parameters which leave invariant the coinvariant together with the mass operator. In this way the theory is formulated in terms of equivalence classes. This is reminiscent of what occurs with the gauge theories, which are defined in terms of equivalence classes, elements on the same class are related by a gauge symmetry. In this sense, we argue that this internal discrete symmetry is the origin of the gauge symmetry in gauged supergravity.

A symplectic torus bundle is defined by $E$ the total space, $F$ the fiber which is the torus of the target-space $T^2$ compact sector and the base space $\Sigma$, which is also a torus. The structure group $G$ corresponds to the group of the symplectomorphism preserving the canonical symplectic two-form on $T^2$. On $\Sigma$, there exists an induced symplectic two-form, obtained from the pullback of the two-form on $T^2$ by the harmonic map from $\Sigma$ to the fiber $T^2$. We notice that the group of symplectomorphisms in $T^2$ or $\Sigma$ is isomorphic to the area preserving diffeomorphisms. The symplectomorphisms in $T^2$ and in $\Sigma$ define the isotopic classes with a group structure $\Pi_0(G)$, in the case under consideration $SL(2,Z)$.

The action of $G$ on the fiber produces an action on the homology and cohomology classes of $T^2$. It reduces to an action of $\Pi_0(G)$. Besides, there is an homomorphism (\ref{monodromy}). Each homomorphism defines a linear representation 
\begin{eqnarray}
     \rho:\Pi_1(\Sigma)\rightarrow SL(2,Z),
\end{eqnarray}
acting on the first homology group in $T^2$, $H_1(T^2)$. Because $H_1(T^2)$ is an abelian group, this homomorphism gives the structure of the $Z(\left[\Pi_1(\Sigma) \right])$-module to each homology and cohomology group on the bundle. Given a monodromy, \cite{Kahn} established the existence of a one-to-one correspondence between the equivalent classes of symplectic torus bundles, induced by the module structure $Z_\rho^2$ on $H_1(T^2)$, and the elements of $H^2(\Sigma, Z_\rho^2)$, the second cohomology group of $\Sigma$ with coefficients $Z_{\rho}^2$. They classify the symplectic torus bundles for a given monodromy in terms of the characteristic class. In the case of a symplectic torus bundle  with base a torus $\Sigma$, the classification in terms of these characteristic classes is equivalent to the classification in terms of coinvariant classes of the monodromy subgroup, acting on ($p,q$) charges. We denote the coinvariant classes simply as coinvariants.

Therefore, the M2-branes on $M_9\times T^2$ with the irreducible wrapping condition, contain two compatible gauge structures. The first one is given by the symplectic (area preserving) structure of the bundle, which ensures the existence of a symplectic connection transforming under symplectomorphisms and it can be extended to a formulation of the M2-brane on a symplectic torus bundle with monodromy, a nontrivial geometrical construction. The second gauge structure is the $U(1)$ principal bundle related to the 2-form flux on $\Sigma$ due to the central charge condition or equivalently due to a nontrivial flux on the target space. Both gauge structures are compatible, and  consequently, they allow the introduction of a twisted torus bundle structure.
\subsection{Symmetries induced by the monodromy on the M2-brane with fluxes.}

Let us consider that the monodromy on the fiber is given by
 \begin{eqnarray}\label{monodromyfiber}
    \mathcal{M}_G=\begin{pmatrix} \mathcal{M}_{11} & \mathcal{M}_{12}\\
                        \mathcal{M}_{21} & \mathcal{M}_{22}
 \end{pmatrix}^{(\alpha+\beta)} \in SL(2,Z),
\end{eqnarray}
where ($\alpha,\beta$) are the integers characterizing the elements of $\Pi_1(\Sigma)$ and specific values of $\mathcal{M}_{ij}$, with $i,j=1,2$ will lead to parabolic, elliptic or hyperbolic monodromies according to it trace. The induced transformation on $\Sigma$, also called induced monodromy on $\Sigma$, is given by
\begin{eqnarray}
   \mathcal{M}_G^*= \Omega^{-1}\mathcal{M}_G(\alpha,\beta)\Omega = \begin{pmatrix} \mathcal{M}_{11} & -\mathcal{M}_{12}\\
                        -\mathcal{M}_{21} & \mathcal{M}_{22}
 \end{pmatrix}^{(\alpha+\beta)} ,\label{inducedmonodromy}
\end{eqnarray}
with $\displaystyle \Omega=\left(\begin{array}{cc}
    -1 & 0  \\
    0 & 1
\end{array}\right)$, equivalently to $S^*_2$ in (\ref{SdualQ}). Therefore symplectomorphisms not connected with the identity on $\Sigma$ are realized by
	\begin{eqnarray}
	  d\widetilde{X}^r \rightarrow (\tilde{g}^*)^r_s d\widehat{X}^s \label{SL(2,Z)sigma11}, \quad
	    \mathbb{W} \rightarrow \mathbb{W}(\tilde{g}^*)^{-1}, \label{SL(2,Z)sigma22} 
	\end{eqnarray}
where $\tilde{g}^*\in\mathcal{M}^*_G$ and the action of $S_U$-duality, when the monodromy is nontrivial, is given by
\begin{eqnarray}
	\tau &\rightarrow& \tau'=\frac{a\tau+b}{c\tau+d} , \label{GSdualtau},
\quad	R\rightarrow R'= R|c\tau+d| , \label{GSdualR} \quad
	A\rightarrow A'=Ae^{i\varphi_\tau},  \quad
	 \\
	\Gamma &\rightarrow& \Gamma'=\Gamma e^{i\varphi_\tau}, \quad W \rightarrow \mathbb{W}'=g^*\mathbb{W},  \label{GSdualW}\quad
	Q \rightarrow Q'=gQ,	\label{GSdualQ}
	\end{eqnarray}
	with $\displaystyle g=\left( \begin{array}{cc}
	 a  & b  \\
	    c & d
	\end{array} \right) \in\mathcal{M}_G$ and $c\tau+d = |c\tau+d|e^{-i\varphi_\tau}$. The M2-brane sectors with central charges are invariant under these $SL(2,Z)$ symmetry transformations on $\Sigma$ and $T^2$.
 
 \subsection{Mass operator}

In order to obtain the mass operator it can be seen that the purely harmonic contributions on the Hamiltonian (\ref{HamiltonianC+}) are given by the same winding and flux term as in  (\ref{massoperatorwinding}). 

 To reproduce the KK term on the mass operator, it can be checked that (\ref{circletocirlce}) with $\mathbb{M}$ given by (\ref{matrixM}) and the winding matrix $\mathbb{W}$, do not reproduce a map onto circles. However, if we consider that $\tilde{l}_8=n\tilde{k}_8$ and $\tilde{m}_8=n\tilde{k}_9$ with $\tilde{k}_8,\tilde{k}_9\in\mathbb{Z}$, the winding matrix can be written as
\begin{eqnarray}
\widetilde{\mathbb{W}}= \mathbb{W}(g^*)^{-1}=\widetilde{S}\left( \begin{array}{cc}
    n & 0  \\
    0 & 1 
\end{array}\right), \label{Swinding}
\end{eqnarray}
with $\displaystyle \widetilde{S}=\left( \begin{array}{cc}
    \tilde{k}_8 & \tilde{l}_9  \\
    \tilde{k}_9 & \tilde{m}_9 
\end{array}\right) \in SL(2,Z)$. Therefore, (\ref{circletocirlce}) can be written as
\begin{eqnarray}\label{mapontocircles83}
\frac{1}{2\pi R}\oint_{C_S}\mathbb{N}^{-1}\left(\begin{array}{c}
    dX^8 \\
    dX^9
\end{array}\right) = \oint _{C_S}\left(\begin{array}{c}
    nd\widetilde{X}^8 \\
    d\widetilde{X}^9
\end{array}\right),
\end{eqnarray}
with $\mathbb{N}^{-1}=\widetilde{S}^{-1}\mathbb{M}^{-1}$, and from (\ref{Conju_Momenta}) we have the KK modes given by $\displaystyle P^0_r = b(\mathbb{N}^{-1})^s_r\frac{\widehat{m}_s}{R}$. Hence
\begin{eqnarray}
P^0_8 = b\frac{t_8}{R},\label{P011} \quad
P^0_9 = b\frac{t_9-t_8\mbox{Re}(\tau)}{R\mbox{Im}(\tau)},\label{P022}
\end{eqnarray}
with
\begin{eqnarray}
t_8 = \tilde{m}_9\widehat{m}_8-\tilde{k}_9\widehat{m}_9, \label{t_1}\quad
t_9 = \tilde{k}_8\widehat{m}_9 - \tilde{l}_9\widehat{m}_8. \label{t_2}
\end{eqnarray}
Consequently, the KK term on the mass operator, can be written in this case as
 \begin{eqnarray}\label{KKterm}
b^2 \frac{\vert t_8\tau-t_9 \vert^2 }{(R\mbox{Im}(\tau))^2}.
\end{eqnarray}
 This term is invariant under (\ref{GSdualtau})-(\ref{GSdualW}). It can be checked that, for a trivial monodromy, the winding matrix can be written as (\ref{Simplifiedwinding}) and we recover the expressions given by (\ref{P01}). Notice that the integers $t_8,t_9$ may be written as $t_8=mq$, $t_9=mp$ with $m\in\mathbb{Z}$ and $p,q$ relatively primes.

 In order to complete the mass operator, we have to also consider the nonzero modes of the Hamiltonian of the M2-brane with $C_\pm$ fluxes and nontrivial monodromy.
 

\section{M2-branes on inequivalent coinvariants with parabolic monodromies}

In this section we present an M2-brane Hamiltonian defined on the inequivalent classes of twisted torus bundles with parabolic monodromies. This is a new result that has not been previously identified in the literature. It corresponds to a functional on the coinvariants associated with a given monodromy. This formulation generalizes the construction in Section 3, which is associated with the particular case of a trivial monodromy. This formulation is defined on the module of $\mathcal{M}_p$-coinvariants. We provide an explicit construction of the model.


\subsection{M2-brane on the module of $\mathcal{M}_p$-coinvariants}

 Inequivalent torus bundles are classified according to the coinvariant classes, briefly coinvariants, for a given monodromy \cite{mpgm3,mpgm2,mpgm7,mpgm10}. The coinvariants related to the fiber and the base manifold are given by 
 \begin{eqnarray}
      C_F &=& \left\lbrace Q + \mathcal{M}_g\widehat{Q}-\widehat{Q}\right\rbrace, \label{C_F}\\
      C_B &=& \left\lbrace W + \mathcal{M}_g^*\widehat{W}-\widehat{W}\right\rbrace, \label{CB}
 \end{eqnarray}
respectively, where $Q = \left( \begin{array}{c}
         p  \\
          q 
     \end{array}\right) $ with $p,q\in\mathbb{Z}$ and $W = \left( \begin{array}{c}
         l_1  \\
          m_1 
     \end{array}\right)$ with $l_1,m_1\in \mathbb{Z}$, $Q$ and $W$ are KK and winding charges, $\widehat{Q}$ and $\widehat{W}$ correspond to arbitrary charges and $\mathcal{M}_g$ is the monodromy subgroup. Given $\mathbb{W}$ as in (\ref{windingmatrix}), we will consider the class of matrices given by 
     \begin{eqnarray}
    \mathbb{W} = \left(\begin{array}{cc}
        l_1  & l_2 \\
         m_1 & m_2
     \end{array} \right)\left(\begin{array}{cc}
        1  & \frac{\lambda}{m} \\
         0 & 1
     \end{array} \right)    
     \end{eqnarray}
     with $\lambda,m \in\mathbb{Z}$ and $l_1=ml_1'$ and $m_1=mm_1'$ with $l_1',m_1'$ relatively primes. These are the most general matrices with $W$ as first column and $det(\mathbb{W})=n$. We use the first column of $\mathbb{W}$ in the definition of $C_B$, but we could have used the second column also. The following reasoning is also valid in both cases. 
     
     If the monodromy is trivial, the coinvariants contain only one element, as discussed in \cite{mpgm7}, but for a nontrivial monodromy class, the coinvariants associated with the base and the fiber contain an equivalence class of KK and winding charges, respectively, related to the same bundle. As shown in \cite{mpgm2}, it is straightforward to see that the Hamiltonian of M2-branes with $C_\pm$ fluxes is invariant in an orbit of charges ($gQ\subset C_F$) generated by the monodromy, with $g\in\mathcal{M}_g$, restricting the  $SL(2,Z)_{T^2}$ transformation (\ref{GSdualtau})-(\ref{GSdualW}). 
     
     Moreover, we will demonstrate that, for the parabolic monodromy, the Hamiltonian is invariant not only in the orbit of charges, but also in the complete coinvariant. It is generated by the abelian parabolic subgroup
\begin{eqnarray}\label{parabolic_monodromy}
     \mathcal{M}_p = \left(\begin{array}{cc}
         1 & 1 \\
         0 & 1
     \end{array}\right) ^{(\alpha+\beta)}.
\end{eqnarray}
This parabolic representation contains the infinite inequivalent conjugate classes of parabolic monodromy
\begin{eqnarray}
   \mathcal{M}_p= \left( \begin{array}{cc}
        1 & k  \\
        0 & 1
    \end{array}\right). \label{parabolicrep}
\end{eqnarray}
The coinvariants (\ref{C_F}) and (\ref{CB}) are given by
\begin{eqnarray}
     C_F &=& \left( \begin{array}{c}
         p +(\alpha+\beta)\widehat{q}  \label{CFP}\\
          q 
     \end{array}\right), \\
     C_B &=& \left( \begin{array}{c}
         l_1 -(\alpha+\beta)\widehat{m}_1  \label{CBP}\\
          m_1 
     \end{array}\right),
\end{eqnarray}
 In this case, $C_F$ and $C_B$ are characterized by $q$ and $m_1$, respectively. Their different values define inequivalent classes of twisted torus bundles with parabolic monodromy.

So far, the mass operator of supermembranes with monodromy contained in $SL(2,Z)$ can be written as
 \begin{eqnarray}
M_{C_\pm}^2 &=&  (TnA_{T^2})^2 + b^2 m^2 \vert \tau_R^T Q\vert^2 +2\widehat{P}_-^0H'^{C_\pm},\label{MassOp_monodromy}
 \end{eqnarray}
with $H'^{C_\pm}$ the hamiltonian of the M2-brane with $C_\pm$ fluxes and nontrivial monodromy and
\begin{eqnarray}
    \tau_R^T = \frac{1}{R\mbox{Im}(\tau)}\left( \begin{array}{cc}
        -1 & \tau 
    \end{array}\right), \quad Q = \left( \begin{array}{c}
         p  \\
         q 
    \end{array} \right)
\end{eqnarray}
This mass operator is invariant under the monodromy $g\in \mathcal{M}_G$. It is consistently defined on the orbit of KK (winding) charges generated by the monodromy $g$ ($g^*$).

For parabolic monodromies, the mass operator of the M2-brane with $C_\pm$ fluxes given by (\ref{MassOp_monodromy})  is invariant if
\begin{eqnarray}
     Q'&=& \Lambda Q, \label{transQ1} \\
     \mathbb{W}' &=& \Lambda^*\mathbb{W}, \label{transW1}\\
       \tau' &=& \tau + \frac{\mathbb{Z}}{q}, \label{transtau1}
\end{eqnarray}
where the $\Lambda$ matrices given by
\begin{eqnarray}
     \Lambda &=& \left( \begin{array}{cc}
        1  & \frac{\mathbb{Z}}{q} \\
         0 & 1
     \end{array} \right),  \label{lambda_SL(2,Q)}\\
     \Lambda^* &=&\Omega^{-1}\Lambda \Omega= \left( \begin{array}{cc}
        1  & -\frac{\mathbb{Z}}{q} \\
         0 & 1
     \end{array} \right), \label{lambdaestrella}
\end{eqnarray}
with $\displaystyle \Omega = \left( \begin{array}{cc}
    -1 & 0 \\
     0 & 1
\end{array}\right)$, define a subgroup of $SL(2,\mathbb{R})$. The transformation (\ref{transQ1}) maps any element of a given coinvariant onto the same coinvariant
\begin{eqnarray}\label{Paraboliccoinv}
     Q&\xrightarrow[]{\Lambda}& C_F=\left(\begin{array}{c}
          \widetilde{\mathbb{Z}} \\
          q
     \end{array}\right),
\end{eqnarray}
with $\widetilde{\mathbb{Z}}=p+\mathbb{Z}$. Furthermore, given two elements of the coinvariant there exists $Z$ such that they are mapped between each other by the corresponding $\Lambda$ in (\ref{lambda_SL(2,Q)}). Together with the transformation of $\tau$, (\ref{transtau1}), leaves invariant KK term
\begin{eqnarray}
     \frac{\vert q'\tau'-p'\vert^2 }{(R\mbox{Im}(\tau'))^2} = 
     \frac{\vert q\tau-p\vert^2 }{(R\mbox{Im}(\tau))^2}, 
\end{eqnarray}

Furthermore, it preserves $dX_h$, the harmonic map on $T^2$ given by (\ref{generalmap}), and thus is a Hamiltonian and full mass operator symmetry. This set of transformations is a generalization of the parabolic monodromy-generated invariance on an orbit of charges ($gQ\subset C_F$).

It can be seen that the matrix $\mathbb{W}'$ is given by
\begin{eqnarray}\label{windingprima}
   \mathbb{W}'&=&  \left(\begin{array}{cc}
        l_1-\frac{\mathbb{Z}}{q}m_1 & l_2-\frac{\mathbb{Z}}{q}m_2  \\
        m_1 & m_2
   \end{array} \right)
\end{eqnarray}
with  $det(\mathbb{W}')=det(\mathbb{W})=n\in \mathbb{Z}$. Although $\mathbb{W}'$ can not be interpreted as a wrapping matrix (unless $m_1$ and $m_2$ are proportional to $q$), it produces the same central charge related to the winding term on the mass operator.

Consequently, the mass operator (\ref{MassOp_monodromy}) is invariant on the equivalence class of charges given by the coinvariant for a given parabolic monodromy.

Furthermore, the charges on the same coinvariant define the same symplectic torus bundle and hence the same physical M2-brane with monodromy. The mass operator of the M2-brane with parabolic monodromy is then expressed in terms of the coinvariant classes of KK charges and winding numbers. We interpret this parabolic symmetry as the origin of the gauge symmetry of the parabolic type II gauge supergravity in 9D.

Let us emphasize that the parabolic coinvariants are characterized by the integer $q$. For each value of $q$ we have a coinvariant, and there are equivalence classes of $p$ and $\tau$ which are related to the same coinvariant.

In particular, if  $m_r=\lambda_r q, \, \lambda_r\in\mathbb{Z}$ for $r=1,2$ we have that
\begin{eqnarray}
      W&\xrightarrow[]{\Lambda^*}& C_B=\left(\begin{array}{c}
         l_1 - \mathbb{Z} \\
          m_1
     \end{array}\right).
\end{eqnarray}
with $\Lambda^*$ given by (\ref{lambdaestrella}), maps any element of a given (base) coinvariant onto the same coinvariant.  In this particular case, $\mathbb{W}'$ is interpreted as a winding matrix and from (\ref{windingprima}), $det(\mathbb{W}')=n$ implies $n$ proportional to $q$. That is, $n$ will, in general, depend on the parabolic coinvariant.

\subsection{Transformations between different coinvariants}
We now consider  a formulation of the twisted parabolic M2-brane in terms of the module of $\mathcal{M}_p$-coinvariants. It follows from the explicit expression of the mass operator that, indeed, it is defined on the coinvariant classes. 
The M2-brane with trivial monodromy, $\mathcal{M}_0=\mathbb{I}$ was analyzed in section 3.

Let us identify the transformations that relate inequivalent classes of M2-brane twisted torus bundles with parabolic monodromy. This is equivalent to determine the transformation which relates the different coinvariant classes associated to $\mathcal{M}_p$. It turns out that this transformation is a symmetry of the formulation. If the monodromy is trivial, each pair of charges ($p,q$) represents a coinvariant and the symmetry of the formulation is $SL(2,Z)$ as determined by \cite{Schwarz6}. For a nontrivial parabolic monodromy, the space of ($p,q$) points is distributed in terms of disjoint coinvariants associated to $\mathcal{M}_p$ and the M2-brane is a theory on the module of $\mathcal{M}_p$-coinvariants.

Firstly, we introduce some formal definitions that will allow us to determine the precise bundle coinvariant transformation. 
Given a group $G$ and a subgroup $H\in G$ we define the following classes
\begin{eqnarray}
    aH=\left\lbrace ah:h\in H \right\rbrace, a\in G, \label{classesequiv}
\end{eqnarray}
There is an equivalence relationship between two elements $a,b\in G$ provided that $b=ah$ for some $h\in H$. This relation can be re-expressed as  $a=bh^{-1}$.

A relevant property is that each element $c\in G$ is contained in one and only one equivalence class. If $c=ah=b\widehat{h}\rightarrow a=b\widehat{h}h^{-1}\in bH$ and then the classes $aH=bH$. Hence $G$ is the disjoint union of the equivalence classes generated by the subgroup $H$.

Given any pair of charges 
$Q=
\left(\begin{array}{c}
     p  \\
     q 
\end{array}\right)
$
with $p,q\in \mathbb{Z}$
and $Q_0=\left(\begin{array}{c}
     1  \\
     0 
\end{array}\right)$, there exists a matrix $V\in GL(2,Z)$, such that
$
Q=VQ_0.
$ It is given by $
 V= \left(\begin{array}{cc}
    p & r  \\
    q & s
\end{array} \right)$ where $r,s$ are not unique.

The most general expression preserving the determinant is
\begin{eqnarray}\label{genexpdet}
    \left(\begin{array}{cc}
    p & r + \lambda p'  \\
    q & s + \lambda q'
\end{array} \right) = \left(\begin{array}{cc}
    p & r  \\
    q & s
\end{array} \right)\left(\begin{array}{cc}
    1 & \frac{\lambda}{m}  \\
    0 & 1
\end{array} \right),
\end{eqnarray}
with $r$ and $s$ unique, $\lambda,m\in\mathbb{Z}$ such that $p=mp'$ and $q=mq'$ with $p',q'$ relatively primes. In fact, if $p(s-\widehat{s})-q(r-\widehat{r}) = 0$, then $(s-\widehat{s})=\frac{q'}{p'}(r-\widehat{r})$ which implies, since the left hand is an integer and $p',q'$ are relatively primes, the existence of $\lambda$ such that
$\widehat{r}-r=p'\lambda$ y $\widehat{s}-s=q'\lambda$. 
Consequently, the most general solution corresponds to 
\begin{eqnarray}
    \widehat{r}=r+\lambda p' ,\quad \widehat{s}=s+\lambda q'.
\end{eqnarray}
\paragraph{Transformation between coinvariants}
Let us define $\mathcal{V}$ as a linear representation of the discrete subgroup $\mathcal{M}_g$. The quotient $\frac{Q}{g\widehat{Q}-\widehat{Q}}$ is the module of $\mathcal{M}_g$-coinvariants \cite{Sharifi}. Two classes $\left\lbrace Q_1 + g\widehat{Q}-\widehat{Q} \right\rbrace$ and $\left\lbrace Q_2 + g\widehat{Q}-\widehat{Q} \right\rbrace$ are disjoint if and only if $Q_1$ and $Q_2$ are not in the same coinvariant. In the case of a parabolic representation -associated to the monodromy of the twisted torus bundle (\ref{parabolicrep})-, the coinvariants
 are given by (\ref{CFP}) and (\ref{CBP}), where $\widehat{Q}$ is an arbitrary element of the space $\mathcal{V}$ and $g$ any element of the subgroup $\mathcal{M}_p$. They are distinguished solely by the value of $q$. 
 
 In order to transform
\begin{eqnarray}
    C_{q_1}\rightarrow C_{q_2},
\end{eqnarray}
we perform the following transformation
\begin{eqnarray}
    C_{q_1}\xrightarrow[]{\Lambda_{q_1}^{-1}} Q_1= \left(\begin{array}{c}
     p_1  \\
     q_1 
\end{array}\right) \rightarrow Q_0=\left( \begin{array}{c}
         1  \\
         0 
    \end{array}\right)\rightarrow Q_2=\left(\begin{array}{c}
     p_2  \\
     q_2 
\end{array}\right) \xrightarrow[]{\Lambda_{q_2}} C_{q_2},
\end{eqnarray}

through $SL(2,Q)$ and $GL(2,Z)$ transformations with $q_1\neq q_2$. 

Following (\ref{classesequiv}), where $G=GL(2,Z)$ and $H=\mathcal{M}_p$, the transformation $Q_1\rightarrow Q_0$ is defined by the equivalence class $a\mathcal{M}_p$  determined by 
\begin{eqnarray}
    a=\left(\begin{array}{cc}
     p_1 & r_1\\
     q_1 & s_1
\end{array}\right),
\end{eqnarray}
where $r_1,s_1$ define $a\in GL(2,Z)$. Consequently, 
\begin{eqnarray}
    \mathcal{M}_pa^{-1}Q_1 = Q_0,
\end{eqnarray}
and 
\begin{eqnarray}
     b \mathcal{M}_pQ_0 = Q_2,
\end{eqnarray}
with $ b=\left(\begin{array}{cc}
         p_2 & r_2  \\
         q_2 & s_2
    \end{array}\right) $. The total transformation is
\begin{eqnarray}
   Q_1 \xrightarrow[]{b \mathcal{M}_pa^{-1}} Q_2,
\end{eqnarray}
or equivalently
\begin{eqnarray}
   Q_1 \xrightarrow[]{ba^{-1}} Q_2.
\end{eqnarray}
Given $p_1,q_1$ ($p_2,q_2$), the most general expression for $a$ ($b$), preserving its determinant has the general expression (\ref{genexpdet}). Notice that $\left(\begin{array}{cc}
     1 & \frac{\lambda}{m} \\
     0 & 1
\end{array}\right)Q_0=Q_0$. Consequently, $a,b$ are determined uniquely by $p_1,q_1$ and $p_2,q_2$ , respectively. 

Each coinvariant $C_q$ contain the element $ \left(\begin{array}{c}
         1  \\
          q_1
    \end{array}\right)$. In this case, we define
$  a= \left(\begin{array}{cc}
        1 & 1  \\
        q_1 & (q_1+1)
    \end{array}\right)$, $b = \left(\begin{array}{cc}
        1 & 1  \\
        q_2 & (q_2+1)
    \end{array}\right),
$
with $q_2\neq q_1$, then it is easy to verify that 
\begin{eqnarray}
    \left(\begin{array}{c}
         1  \\
          q_2
    \end{array}\right) =ba^{-1}  \left(\begin{array}{c}
         1  \\
          q_1
    \end{array}\right)
\end{eqnarray}
with 
\begin{eqnarray}
    \mathcal{M}_\beta \equiv ba^{-1} = \left(\begin{array}{cc}
        1 & 0  \\
        \beta & 1
    \end{array}\right).
\end{eqnarray}
The elements of $\mathcal{M}_\beta$ determine a group conjugate to $\mathcal{M}_p$. The transformation between coinvariants is given by
\begin{eqnarray}
    C_{q_1}\xrightarrow[]{\Lambda_{q_1}^{-1}} \left(\begin{array}{c}
     1  \\
     q_1 
\end{array}\right) \xrightarrow[]{\mathcal{M}_\beta}  \left(\begin{array}{c}
     1  \\
     q_2 
\end{array}\right) \xrightarrow[]{\Lambda_{q_2}} C_{q_2}.
\end{eqnarray}
Let us emphasize that this transformation maps integer charges into integer charges. Moreover, as the $SL(2,\mathbb{Q})$ transformations are within the coinvariant, is the parabolic matrix $\mathcal{M}_\beta$ the one that characterize the transformation between coinvariants.

 Consequently, there is a transformation
\begin{eqnarray}
	  C_{q_{1}} &\xrightarrow[]{\widetilde{\Lambda}}&  C_{q_{2}},\quad \tau \rightarrow \frac{\left( 1 + \frac{\mathbb{Z}_2}{q_2}\beta\right)\tau + \left(-\frac{\mathbb{Z}_1}{q_1} + \frac{\mathbb{Z}_2}{q_2}\left(1-\frac{\mathbb{Z}_1}{q_1}\beta\right)\right)}{\beta\tau+1- \frac{\mathbb{Z}_1}{q_1}\beta}, 
\quad \mathbb{W}\rightarrow \widetilde{\Lambda}^* \mathbb{W}  , \nonumber \\
	 R&\rightarrow& R\vert \beta\tau+1- \frac{\mathbb{Z}_1}{q_1}\beta\vert, \quad A\rightarrow Ae^{i\varphi_\tau},  \quad
	\Gamma \rightarrow \Gamma e^{i\varphi_\tau}, \quad \label{115}
	\end{eqnarray}
	with 
	\begin{eqnarray}
	   \widetilde{\Lambda}&=& \Lambda_{q_2}\mathcal{M}_\beta \Lambda_{q_1}^{-1} = \left(\begin{array}{cc}
         1 + \frac{\mathbb{Z}_2}{q_2}\beta & -\frac{\mathbb{Z}_1}{q_1} + \frac{\mathbb{Z}_2}{q_2}\left(1-\frac{\mathbb{Z}_1}{q_1}\beta\right)  \\
        \beta & 1- \frac{\mathbb{Z}_1}{q_1}\beta
    \end{array}\right) , \label{lambdatilde} \\
    \Lambda^* &=& \Omega^{-1}\Lambda\Omega
	\end{eqnarray}
	and $e^\varphi_\tau=\frac{\beta\tau+1- \frac{\mathbb{Z}_1}{q_1}\beta}{|\beta\tau+1- \frac{\mathbb{Z}_1}{q_1}\beta|}$ leaving invariant the M2-brane mass operator.
	
	This can be interpreted as a duality between inequivalent classes of M2-brane twisted torus bundles with parabolic monodromies.
		
	One could also use the lower triangular parabolic matrix to describe the parabolic monodromy, and then an upper triangular parabolic matrix  describes the transformation between the parabolic coinvariants. Since both matrices are in the same conjugacy class, the M2-brane mass operator also remains invariant in this case.

\section{\textit{Parabolic} ($p,q$)-strings.}
 The identification of type IIA with 11D supergravity on a circle and T-duality between type II theories induces the relation of type IIB on a circle with 11D supergravity on a torus. We have seen in (\ref{sec3}) that M2-branes on $M_9\times T^2$ with irreducible wrapping yield type $IIB$-string compactified on a circle.

We will now show that the double-dimensional reduction of M2-branes with $C_{\pm}$ fluxes and parabolic monodromy is related to ($p, q$)-superstrings compactified on a circle, with the corresponding restriction on the $SL(2, Z) $ symmetry provided by the monodromy.

In \cite{mpgm2}, the low energy limit of M2-brane with monodromy contained in the conjugacy classes of $SL(2,Z)$, i.e. parabolic, elliptic and hiperbolic according to it trace\footnote{In this paper we will not discuss the case in which the monodromy is nonlinearly realized.}, were related to the type IIB gauged supergravity with parabolic, elliptic and hyperbolic monodromy, respectively. In the conclusion we will discuss the relation of the parabolic string with the uplift of type IIB parabolic supergravity in 9D.
\subsection{Mass operator of the \textit{parabolic} ($p,q$)-string}
The mass operator of the M2-brane with $C_\pm$ fluxes and nontrivial monodromy (\ref{MassOp_monodromy}) is defined in the orbit of charges for a given monodromy $g\in\mathcal{M}_p$  (\ref{GSdualtau})-(\ref{GSdualW}). For parabolic monodromies, we have shown that it can be consistently formulated on the coinvariants (\ref{transQ1})-(\ref{transtau1}) which classify inequivalent twisted torus bundles.

In order to obtain the full mass operator and the corresponding ($p,q$)-strings, we consider the string configurations on the $H_{C_\pm}\vert_{SC}$ Hamiltonian on (\ref{MassOp_monodromy}) as in section 3 but with the harmonic map given by
\begin{eqnarray}\label{harmonicmaptheta}
      dX_h = 2\pi R (l_r+m_r\tau)\Theta^r_sd\widehat{X}^s,
 \end{eqnarray}
with $\Theta^r_s=\delta^r_s$ according to the transformation (\ref{SL(2,Z)sigma22}). It lead us to the Hamiltonian given by (\ref{hamiltonianSC}), with the bosonic and fermionic potential written as (\ref{Bosonicpotential}) and (\ref{fermionicpotential}), respectively. The harmonic map (\ref{harmonicmaptheta}) can be used if we consider
	\begin{eqnarray}
	  d\widetilde{X}^r \rightarrow S^r_s d\widehat{X}^s, \quad
	    \mathbb{W} \rightarrow \mathbb{W}(\tilde{g}^*)^{-1}, \quad \widetilde{\Theta} \rightarrow g^*\Theta S^{-1} \label{SL(2,Z)sigma222} 
	\end{eqnarray}
with $S$ and arbitrary matrix of $SL(2,Z)$, instead of (\ref{SL(2,Z)sigma22}). The matrix $\Theta$ will in general depend on the monodromy.

Let us perform the same change on the canonical basis of homology and the corresponding basis of harmonic one-forms as in (\ref{homologybasis1}). Nevertheless, instead of using the full $SL(2,Z)_{T^2}$ and $SL(2,Z)_\Sigma$, we will only use the  restricted $SL(2,Z)_\Sigma$ symmetry given  by the induced monodromy in (\ref{SL(2,Z)sigma11}). In this case, the Hamiltonian remains invariant under such transformation, but the winding matrix transform according to (\ref{SL(2,Z)sigma22}). It is evident that this transformation leaves invariant the harmonic one-form. 

Therefore, the Hamiltonian of the M2-brane in the string configurations with nontrivial monodromy is given by
\begin{eqnarray}
H_{C_-}\vert_{SC} &=& \int d\widetilde{X}^8\wedge d\widetilde{X}^9 \left\lbrace \frac{1}{2P_-^0} \left(\frac{P'_M}{\sqrt{w}}\right)^2 + \frac{T^2}{2P_-^0} (2\pi R \vert l_9+m_9\tau \vert)^2\partial_8X^M \partial^8X_M  \right.  \nonumber \\
&-& \left. T(2\pi R)\Bar{\theta}\Gamma^-\left[l_9\Gamma_8 + m_9(\Gamma_8 \mbox{Re}(\tau)+\Gamma_9 \mbox{Im}(\tau))\right] \partial_8\theta \right\rbrace.
\end{eqnarray}
Is easy to see that following the proposition III.2.3 from \cite{Farkas} we can rewrite the Hamiltonian as
\begin{eqnarray}
H_{C_-}\vert_{SC} &=& \frac{1}{2P_-^0}\int d\widetilde{X}^8 \left\lbrace \left(\frac{P'_M}{\sqrt{w}}\right)^2 + T^2(2\pi R \vert l_9+m_9\tau \vert)^2\partial_8X^M \partial^8X_M  \right.  \nonumber \\
&-& \left. 2P^0_- T(2\pi R)\Bar{\theta}\Gamma^-\left[ l_9\Gamma_8 + m_9(\Gamma_8 \mbox{Re}(\tau)+\Gamma_9 \mbox{Im}(\tau)) \right] \partial_8\theta \right\rbrace.
\end{eqnarray}
If we consider the global constraint as in the previous section, we have that the one corresponding to $\widetilde{X}^9$ lead us to
\begin{eqnarray}
    0 = t_8l_9 + t_9m_9,
\end{eqnarray}
from where we obtain that $\widehat{m}_9=0$ as in the previous section and therefore 
$t_8 = m_9\widehat{m}_8$, $t_9 = -l_9 \widehat{m}_8$. Before analyzing the global constraint for $\widetilde{X}^8$, let us recall that $\widetilde{X}^8$ is adimensional. Therefore, we consider $\xi$ given by (\ref{sigmadimensions}) but with $a=\frac{K\sqrt{\widehat{P}_-^0}}{\widetilde{T}}$ with $\widetilde{T}=T(2\pi R)\vert l_9+m_9\tau \vert$ and $K$ a constant with dimensions of $(energy)^{1/2}$. In consequence, the Hamiltonian can be written as
\begin{equation}
\small H_{C_-}\vert_{SC} = \frac{K}{\sqrt{P_-^0}}\int d\xi\left\lbrace \frac{1}{2\widetilde{T}}
\left(\frac{P^*_M}{\sqrt{w}}\right)^2  +  \frac{\widetilde{T}}{2}\partial_\xi X^M \partial^\xi X_M - \frac{\bar{S^*}}{\sqrt{w}}\Gamma^* \partial_\xi\theta \right\rbrace ,
\end{equation}
where $M=1,\dots,8$ and 
\begin{eqnarray}
   \Gamma^* = \frac{(l_9 + m_9\mbox{Re}(\tau))\Gamma_8 + m_9\mbox{Im}(\tau)\Gamma_9}{\vert l_9 + m_9\tau \vert},
\end{eqnarray}
satisfies the corresponding Clifford algebras.

Following the same decomposition as in the previous section, we can write the string Hamiltonian in terms of the re-scaled $SO(7)$ spinors (\ref{SO(7)reescaled}) as
\begin{equation}\label{Hamiltoniantrivialmonodromy2}
H_{C_-}\vert_{SC} = \int d\xi \left\lbrace \frac{1}{2\widetilde{T}}
\left(\frac{P'_M}{\sqrt{w}}\right)^2 + \frac{\widetilde{T}}{2}\partial_\xi X^M \partial^\xi X_M - \frac{i}{\sqrt{2}}\frac{P_-^0}{a} (\widehat{\lambda}^1\partial_\xi \widehat{\lambda}^1 - \widehat{\lambda}^2\partial_\xi \widehat{\lambda}^2)\right\rbrace, 
\end{equation}
and the same expression in terms of the oscillators is given by
\begin{eqnarray}
H_{C_-}\vert_{SC} &=&   T8\pi^2 R\vert m_9 \tau +l_9 \vert (N_T + \bar{N}_T),
\end{eqnarray}
where we have set $c=\hbar=1$.

Consequently, the mass operator corresponding to an M2-brane with $C_-$ fluxes and monodromy, can be written on the strings confiurations as
\begin{equation}
M_{C_-}^2 =  (TnA_{T^2})^2 +  \frac{\widehat{m}_8\vert (m_9 \tau +l_9) \vert^2 }{(R\mbox{Im}(\tau))^2}
+T8\pi^2 R\vert m_9 \tau +l_9 \vert (N_T + \bar{N}_T), \label{MassOp_monodromy2}
 \end{equation}
and it remains invariant under the transformations given by (\ref{SL(2,Z)sigma11}), (\ref{GSdualtau}) and (\ref{GSdualW}), respectively, restricted by the monodromy $g\in\mathcal{M}_g$.

Finally, from the mass operator (\ref{MassOp_monodromy2}) we notice that
\begin{eqnarray}
    M^2_{C_{-}} &=&  (T_{11}nA_{T^2})^2 + \left( \frac{\widehat{m}_8m_1\vert q\tau-p\vert}{R \mbox{Im}(\tau)}\right)^2 +  T8\pi^2 Rm_1\vert q\tau-p \vert(N_T + \bar{N}_T),   \nonumber \\
    &=& (T_{11}nA_{T^2})^2 + \widehat{m}_8^2\vert \tau_R^T Q\vert^2 + T8\pi^2 R^2 \mbox{Im}(\tau)\vert \tau_R^T Q\vert (N_T + \bar{N}_T), \\ 
    M^2_{C_{+}} &=& M^2_{C_{-}} - 2P_-^0 TA_{T^2}nk_{+}, \label{MassOperatorFinal44}
\end{eqnarray}
with 
\begin{eqnarray}
    m_1q=m_9,  \quad
    m_1p=-l_9, \\
    \tau_R^T = \frac{m_1}{R\mbox{Im}(\tau)}\left( \begin{array}{cc}
        -1 & \tau 
    \end{array}\right), \quad Q = \left( \begin{array}{c}
         p  \\
         q 
    \end{array} \right)
\end{eqnarray}
and it is consistently defined on the orbit of KK charges generated by any monodromy $g\in\mathcal{M}_g$.

As happens in (\ref{MassOp_monodromy}) for the M2-brane, the mass operator (\ref{MassOperatorFinal44}) is invariant on the coinvariants for a given parabolic monodromy $g\in\mathcal{M}_p$, hence describing the same  twisted torus bundle with parabolic monodromy  description. Inequivalent coinvariants are given by different values of $q'=m_1q$, while $p'=m_1p\in\mathbb{Z}$ defines the different elements within the same class. At string theory level these coinvariant classes defines the equivalence classes of charges.

If we now follow the same procedure as in section 3, we can obtain the parabolic $(p,q)$-string mass operator given by  
\begin{eqnarray}\label{cuerda IIBGauge}
 M^2_{C_q} = \left( \frac{n}{R_B} \right)^2 + (2\pi R_B \widehat{m}_8 T_{C_q})^2 + 4\pi T_{C_q}(N_L + N_R) - \frac{2P_-^0T_{C_q}^{1/6}R_B^{-2/3}nk_+}{\vert \widehat{\lambda}^T C_q \vert^{1/6}},
\end{eqnarray}
where  
\begin{eqnarray}
\tau=\lambda_0,\quad \beta^2 = \frac{TA_{T^2}^{1/2}}{T_c}, \quad
 R_B^2 = (TA_{T^2}^{3/2}T_c)^{-1}, \label{tau,beta_y_R2},
 \end{eqnarray}
 and
\begin{eqnarray}
    T_{C_q} &\equiv& \vert \widehat{\lambda}^T C_{q}\vert T_c, \quad \widehat{\lambda}^T = \frac{m_1}{(\mbox{Im}(\lambda_0))^{1/2}} \left( \begin{array}{cc}
    -1 & \lambda_0
\end{array} \right)
\end{eqnarray}
with $C_q$ as in (\ref{Paraboliccoinv}), $T_c = T^{2/3}$ the string tension as in (\ref{tau,beta_y_R}) and $\lambda_0 = \xi + i\exp{\phi_{0}}$ the axion-dilaton of the type IIB theory.

The associated pair of $(p,q)$ charges of the parabolic string, with $p,q$ relatively primes, gets all identified for any given $q$ from an 11D point of view.

In fact, from (\ref{transQ1})-(\ref{transtau1}), we have that the mass operator (\ref{cuerda IIBGauge}) is invariant under the following $SL(2,\mathbb{Q})$ transformation
\begin{eqnarray}
     Q'&=& \Lambda Q, \label{transQ1string} \\
     \lambda_0' &=& \lambda_0 + \frac{\mathbb{Z}}{q}, \label{transaxiondilaton}
\end{eqnarray}
where $\Lambda$ is given by (\ref{lambda_SL(2,Q)}). This symmetry relates states of parabolic ($p,q$)-strings, with $q$ fixed and $p$ any coprime number, with the same local and global origin in 11D. The tension $T_{C_q}$ is invariant under this transformation. This transformation is residual from the one obtained in (\ref{transQ1})-(\ref{transtau1}). It defines inequivalent classes of parabolic ($p,q$)-strings compactified on a circle.

The transformation that relates different values of $q$, is given by $\widetilde{\Lambda}$ according to (\ref{lambdatilde})
\begin{eqnarray}
   C_{q_2} = \widetilde{\Lambda} C_{q_1}.
\end{eqnarray}
Therefore, it can be checked from (\ref{115}), that
\begin{eqnarray}
    Q' &=& \widetilde{\Lambda} Q, \nonumber \\
    \lambda_0' &=& \frac{\left( 1 + \frac{\mathbb{Z}_2}{q_2}\beta\right)\lambda_0 + \left(-\frac{\mathbb{Z}_1}{q_1} + \frac{\mathbb{Z}_2}{q_2}\left(1-\frac{\mathbb{Z}_1}{q_1}\beta\right)\right)}{\beta\lambda_0+1- \frac{\mathbb{Z}_1}{q_1}\beta}
\end{eqnarray} 
leaves invariant the tension $T_{C_q}$ and hence the parabolic string mass operator (\ref{cuerda IIBGauge}).

As a result, we will have parabolic ($p,q$)-strings on $M_9\times S^1$, which are obtained through a double dimensional reduction from M2-branes with parabolic monodromy. Moreover, the M2-branes low energy limit in 9D is related to type IIB gauged supergravities in 9D. In fact, it was already shown in \cite{mpgm2} that the eight inequivalent classes of M2-branes with nontrivial monodromy are in correspondence, in the low energy limit, with the type II gauged supergravities in 9D. Therefore, we claim that the corresponding parabolic ($p,q$)-string must be associated with the type IIB gauged supergravity in 9D with a parabolic gauging group.

\section{Conclusions}
We characterize the string description of the toroidally wrapped M2-branes with KK charges and a quantized three-form $C_3$, that induces two-form fluxes on the target formulated on a twisted torus bundle with monodromy. In the case of a trivial monodromy the formulation reduces exactly to the one considered in \cite{Schwarz6} and the double dimensional reduction coincides with the ($p,q$)-strings with $SL(2,Z)$ symmetry. We analyze the formulation for nontrivial parabolic monodromies and identify the "gauge" symmetry in the M2-brane formulation related to the associated supergravity. We perform explicitly the construction for a generic parabolic monodromy, we discuss the other monodromies elsewhere.

We first characterize the role of the central charge in the supermembrane double reduction on a Minkowski target space toroidally wrapped. It is well-known that a supermembrane on a torus is associated to a wrapped type IIB $SL(2,Z)$ $(p,q)$ string on a circle \cite{Schwarz6}. M2-brane with $C_\pm$ fluxes has a purely discrete mass spectrum. It is equivalent, through a canonical transformation of the phase space variables, to the  M2-brane with central charge \cite{Restuccia}. The equivalence is exact when only  $C_-$ fluxes  are present and  the $C_+$ component vanishes \cite{mpgm6} and it has a constant shift in the presence of  $C_+$ fluxes \cite{mpgm6}. We show that the existence of a central charge condition is a  necessary prerequisite to obtain the sectors of the  $(p,q)$ string mass operator with $p,q \ne 0$, which are associated with string bound states. A central charge condition is necessary to define the embedding map  onto circles and hence, an actual wrapping of the M2-brane on a torus. The type IIB ($p,q$)-string KK-term is inherited from the central charge condition. Furthermore the characteristic tension of the wrapped $(p,q)$ string with $p,q \ne 0$  cannot be obtained from vanishing central charge (reducible wrapping) in the M2-brane theory.

We obtain by double dimensional reduction the $(p,q)$ string associated with the  M2-brane with $(C_\pm)$ fluxes. It inherits a new constant topological term We find that it inherits one constant extra topological term associated to the amount of flux $C_+$ turned on. We analyze three different classes of M2-brane twisted torus bundles, attending to the values of the monodromy:

When the monodromy is trivial,  the symmetries on $T^2$ and $\Sigma$ are given by the full group $SL(2,Z)$ and for $C_-\neq 0$ and $C_+=0$ the results coincide with those obtained by \cite{Schwarz6}. The coinvariant class contains solely one element, $Q$. Different coinvariants are related by an $SL(2,Z)$ transformations. When doubled dimensionally reduced, each  wrapped $(p,q)$ strings is connected by an $SL(2,Z)$ transformation.
 
We have concentrated this study on the case when the M2-brane is formulated on a twisted torus bundle with monodromy contained in $SL(2,Z)$, the group of isotopy classes of area preserving diffeomorphisms (symplectomorphisms) \cite{mpgm2,mpgm3}. In that case, the discrete symmetry is restricted by the inequivalent classes of the monodromy subgroups, generated by elliptic, parabolic, and hyperbolic $SL(2,Z)$ matrices. The inequivalent classes of twisted torus bundles are given by the coinvariants on the fiber and base manifold, for a given monodromy $\mathcal{M}_g$ and $C_{\pm}$ flux. 
The Hamiltonian of the M2-brane with $C_\pm$ fluxes is invariant on an orbit of charges $gQ\subset C_F$ generated by $g\in\mathcal{M}_g$ in \cite{mpgm2}. We show here, that the Hamiltonian with parabolic monodromies, can be consistently defined on the coinvariant $C_F$. There are infinite inequivalent coinvariants associated to the parabolic monodromy $\mathcal{M}_p$. They are determined by one of the KK charges, hence they are classified by the integers. The mass operator's symmetry group is an extension of the subgroup generated by a parabolic generator in $SL(2,Z)$. Its generator is a parabolic matrix in $SL(2,\mathbb{Q})$, $\mathbb{Q}$ being the rational numbers. This symmetry is not present in \cite{Schwarz6} where each coinvariant has solely one element. When the monodromy is nontrivial, we identify the symmetry relating the elements of the coinvariant as a "gauge symmetry" of the formulation. In fact, not only the physical content remains invariant but also the geometric formulation is defined on the same twisted torus bundle.

 We demonstrate that the transformation between M2-brane twisted torus bundles with parabolic monodromy but different second cohomology class, i.e. different coinvariants, can be expressed in terms of a subgroup $\mathcal{M}_{\beta}$ conjugated to the $\mathcal{M}_p$. It leaves  the  M2-brane mass operator invariant, although they describe formulations of M2-brane on inequivalent symplectic torus bundles.

These sectors, globally  described in terms of twisted torus bundles with nontrivial monodromy, are described  at low energies by the Type II gauged supergravities in 9D. The double dimensional reduction of the M2-brane Hamiltonian yields a Hamiltonian of a class of ($p,q$)-string with a parabolic $SL(2,\mathbb{Q})$ symmetry, inherited from the monodromy of the M2-brane from which it descends.  Therefore, we  obtain  a class of ($p,q$)-strings given by the parabolic conjugacy classes of the monodromy. These ($p,q$)-strings have an origin in 11D on the nontrivial sectors of M2-brane described by the inequivalent classes of twisted torus bundles with parabolic monodromy.  Their low energy must be the same that the nontrivial M2-branes, i.e. the type IIB gauged supegravities in 9D. These \textit{parabolic} $(p,q)$-\textit{strings} may correspond to the parabolic Scherk-Schwarz reduction of type IIB superstring, considered in \cite{Hull8} in terms of F-theory compactified on a twisted torus.

\appendix

\section*{Appendix A}
Let us recall that $M=1,\dots,8$ on the Hamiltonian (\ref{Happendix1}). Moreover, as we are considering string configurations, we may compare this expression with the string type II Hamiltonian in the closed sector. 

Let us consider next representation of gamma matrices in 11 dimensions
\begin{eqnarray}
\Gamma^{+}= i\sqrt{2}\begin{pmatrix} 0 & \mathbb{I}_{16\times 16} \\
                       0 & 0 
\end{pmatrix} \label{Gammamas},
\Gamma^{-}= i\sqrt{2} \begin{pmatrix} 0 & 0 \\
            -\mathbb{I}_{16\times 16} & 0 
\end{pmatrix} \label{Gammamenos},
\Gamma^{a}=\begin{pmatrix} \gamma^a & 0\\
            0 &  -\gamma^a
\end{pmatrix} \label{Gammaa}
\end{eqnarray}
where $\gamma^a\in SO(9)$ are $16\times 16$ matrices and $a=(m,r)$ with $m=1,\dots,7$ and $r=8,9$. It can be check that these representations satisfies the anticommutation relations 
\begin{eqnarray}
\left\lbrace\Gamma^+,\Gamma^- \right\rbrace = 2\mathbb{I}_{32} , \label{Clifford1} \quad
\left\lbrace \Gamma^{\pm} , \Gamma^a \right\rbrace = 0, \label{Clifford2}\quad
\left\lbrace \Gamma^{a} , \Gamma^b \right\rbrace = 2\eta^{ab} \label{Clifford3}
\end{eqnarray}
 We may choose
\begin{eqnarray}
\theta = \left(\begin{array}{c}
     \psi  \\
     0
\end{array}\right), \label{spinor1} \quad
\bar{\theta} =  \left(
\begin{array}{cc}
     0 & -\psi^T
\end{array} \right), \label{spinor2}
\end{eqnarray}
such that $\Gamma^+\theta=0$. Therefore, in terms of the $SO(9)$ Majorana spinor $\psi$, it can be seen that the fermionic term is given by
\begin{eqnarray}
\frac{\bar{S}^*}{\sqrt{W}}\Gamma^*\partial_\xi\theta &=& i \sqrt{2} P_-^0 \psi^T\gamma_8\partial_\xi \psi
\end{eqnarray}
where the representation of $SO(9)$ matrices is given by 
\begin{eqnarray}
    \gamma_1 &=& -\sigma_2 \otimes \sigma_2 \otimes \sigma_2 \otimes \sigma_1, \label{gamma1} \\
    \gamma_2 &=& -\sigma_2 \otimes \sigma_2 \otimes \sigma_2 \otimes \sigma_2, \label{gamma2}\\
    \gamma_3 &=& -\sigma_2 \otimes \sigma_2 \otimes \sigma_2 \otimes \sigma_3, \label{gamma3}\\
    \gamma_4 &=& \sigma_2 \otimes \sigma_2 \otimes \sigma_1 \otimes \mathbb{I}, \label{gamma4}\\
    \gamma_5 &=& \sigma_2 \otimes \sigma_2 \otimes \sigma_3 \otimes \mathbb{I}, \label{gamma5} \\
    \gamma_6 &=& -\sigma_2 \otimes \sigma_1 \otimes \mathbb{I} \otimes \mathbb{I}, \label{gamma6}\\
    \gamma_7 &=& -\sigma_2 \otimes \sigma_3 \otimes \mathbb{I} \otimes \mathbb{I}, \label{gamma7}\\
    \gamma_8 &=& \sigma_1 \otimes \mathbb{I} \otimes \mathbb{I} \otimes \mathbb{I}, \label{gamma8}\\
    \gamma_9 &=& \sigma_3 \otimes \mathbb{I} \otimes \mathbb{I} \otimes \mathbb{I}, \label{gamma9}
\end{eqnarray}
with
\begin{eqnarray}
    \sigma_1 = \left[\begin{array}{cc}
        0 & 1 \\
        1 & 0
    \end{array} \right], \quad \sigma_2 = \left[\begin{array}{cc}
        0 & i \\
        -i & 0
    \end{array} \right] , \quad \sigma_3 = \left[\begin{array}{cc}
        1 & 0 \\
        0 & -1
    \end{array} \right]
\end{eqnarray}
the Pauli Matrices $2\times 2$ and $\sigma_0=\mathbb{I}$. It can be seen that the $SO(9)$ spinor can be splitted as $\psi=\psi^+ + \psi^-$ with
\begin{eqnarray}
   \psi^+ = P_+\psi = \left( \begin{array}{c}
        \chi^+  \\
        0
   \end{array} \right), \quad
   \psi^- = P_-\psi = \left( \begin{array}{c}
        0  \\
        \chi^-
   \end{array} \right)
\end{eqnarray}
and $P_\pm = \frac{1}{2}(\mathbb{I} \pm \gamma_9)$ written in terms of the chiral matrix of $SO(8)$ such that
\begin{eqnarray}\label{weyleq}
    \gamma_9 \psi^\pm = \pm \psi^\pm.
\end{eqnarray}

\section*{Appendix B}

The Hamiltonian (\ref{Hamiltoniantrivialmonodromy1}), or equivalently (\ref{Hamiltoniantrivialmonodromy2}), are reminiscent of the LCG type II superstring Hamiltonian. It can be checked that the equations of motion for the bosonic variables are given by 
\begin{eqnarray}\label{bosoniceq}
\partial_t^2 X^M = c^2\partial_\xi^2 X^M,
\end{eqnarray}
if $K=\sqrt{P_-^0}$. On the other hand, the equations of motion for the fermionic variables are given by the standard expressions
\begin{eqnarray}
    (\partial_\tau + c\partial_\xi)\widehat{\lambda}^1 = 0, \label{fermioniceq2} \quad
    (\partial_\tau - c\partial_\xi)\widehat{\lambda}^2 = 0, \label{fermioniceq1} 
\end{eqnarray}
In order to obtain the mass operator in terms of the oscillators, let us impose the boundary conditions on the bosonic and fermionic fields. The canonical pairs of bosonic variables is given by  $(X^m,P_m)$ y $(X^*,P_*)$. Therefore, the periodic boundary conditions characteristic of closed strings is 
\begin{eqnarray}
X^m (\xi + a,\sigma^0)&=&X^m (\xi,\sigma^0), \\
X^* (\xi + a,\sigma^0) &=& X^* (\xi,\sigma^0) + (2\pi R_B)\widehat{n},
\end{eqnarray}
However, we know that the Hamiltonian (\ref{Hamiltoniantrivialmonodromy1}) corresponds to the excitations of the nontrivial M2-branes with respect to the center of mass. The zero modes contributions has been used on the winding and KK term. In consequence, we have that
\begin{eqnarray}\label{BosonicBC88}
X^M (\xi,\sigma^0) = - i\sqrt{\frac{\alpha'}{2}}\sum_{n\in\mathbb{Z}-\left\lbrace 0 \right\rbrace} \left[ \frac{\alpha_n^M}{n}e^{\frac{2i\pi n(\xi+\sigma^0)}{a}} + \frac{\widetilde{\alpha}_n^M}{n}e^{\frac{-2i\pi n(\xi-\sigma^0)}{a}}  \right],
\end{eqnarray}
with $M=1,\dots,8$. If we set $c=\hbar=1$ we have the M2-brane tension has dimensions of $\displaystyle \frac{1}{\left[L\right]^3}$ and the string tensions can be written as $\displaystyle T_c = \frac{1}{2\pi \alpha'}$ with the fundamental length given by $l=\sqrt{2\alpha'}$. Therefore, as $a=2\pi l$ we have that
$\displaystyle
    \frac{\alpha'}{2} = \frac{a}{4\pi \widehat{P}_-^0}.
$
 On the GS formalism, the boundary conditions on the spinors are given by
\begin{eqnarray}
 \widehat{\lambda}^1(\xi,\tau) = \widehat{\lambda}^1(\xi+a,\tau), \quad
\widehat{\lambda}^2(\xi,\tau) = \widehat{\lambda}^2(\xi+a,\tau),
\end{eqnarray}
then
\begin{eqnarray} \label{FermionicBC}
\widehat{\lambda}^1&= \sum_n \beta_n^1 e^{\frac{i2\pi n(\tau+\xi)}{l}}, \quad
\widehat{\lambda}^2 = \sum_n \beta_n^2 e^{\frac{-i2\pi n(-\tau+\xi)}{l}}.
\end{eqnarray}
Finally, inserting this on the Hamiltonian (\ref{Hamiltoniantrivialmonodromy1}) and using the standard (anti)-commutation brackets for the (fermionic) bosonic oscillators we have that
\begin{eqnarray}
    H_{C_-}\vert_{SC} &=& T 8\pi^2 R'\vert\tau'\vert (N_T + \bar{N}_T),
\end{eqnarray}
where
$
    N_T = N_B + N_F^1, \label{TotalNObos}\quad
    \bar{N}_T = \bar{N}_B + N_F^2, \label{TotalNOfer}
$ 
are the total number operators 
\begin{eqnarray}
   \sum_{n\neq 0} \alpha_{n}^M\alpha_{-n}^M = 2(N_B + E_0), \\
   \sum_{n\neq 0} \widetilde{\alpha}_{n}^M\widetilde{\alpha}_{-n}^M = 2(\bar{N}_B + \bar{E}_0), \\
   \sum_{n\neq 0} n\beta_{n}^1\beta_{-n}^1 = -2(N_F^1 - \bar{E}_0), \\
   \sum_{n\neq 0} n\beta_{n}^2\beta_{-n}^2 = -2(N_F^2 - E_0)\, .
\end{eqnarray}
and the vacuum energies has been cancelled as in the Ramond sector on the NSR formalism.

\acknowledgments

MPGM thanks to the IFT(UAM-CSIC) for kind hospitality while part of this work was done. CLH is supported by CONICYT PFCHA/DOCTORADO BECAS CHILE/2019-21190263, and the Projects MINEDUC-UA, ANT1956 and MINEDUC-UA, ANT2156. MPGM and CLH also thanks to SEM 18-02 funding project from U. Antofagasta, and to the international ICTP Network NT08 for kind support. 



\providecommand{\href}[2]{#2}\begingroup\raggedright
\endgroup

\end{document}